\documentclass[fleqn,usenatbib]{mnras}

\usepackage{xtab, afterpage}
\usepackage{mathptmx}

\usepackage[T1]{fontenc}

\DeclareRobustCommand{\VAN}[3]{#2}
\let\VANthebibliography\thebibliography
\def\thebibliography{\DeclareRobustCommand{\VAN}[3]{##3}\VANthebibliography}

\usepackage{graphicx}	
\usepackage{amsmath}	
\usepackage{amssymb}	
\usepackage{xspace}
\usepackage{longtable}
\usepackage{subfig}
\usepackage{booktabs}
\usepackage{lscape}
\usepackage{siunitx}
\usepackage[symbol]{footmisc}
\usepackage[dvipsnames]{xcolor}
\usepackage{multirow}

\newcommand{\tls}{\textsc{tls}\,}
\newcommand{\scipy}{\textsc{scipy}\,}
\newcommand{\bls}{\textsc{bls}\,}
\newcommand{\wotan}{{\fontfamily{pcr}\selectfont wötan}\,}
\newcommand{\python}{\textsc{python}\,}
\newcommand{\astropy}{\textsc{astropy}\,}

\newcommand{\tpfplotter}{\textsc{tpfplotter}\,}
\newcommand{\vespa}{\textsc{vespa}\,}
\newcommand{\pastis}{\textsc{pastis}\,}
\newcommand{\felix}{\textsc{felix}\,}
\newcommand{\vetter}{\textsc{vetter}\,}
\newcommand{\dave}{\textsc{dave}\,}
\newcommand{\robovetter}{\textsc{robovetter}\,}
\newcommand{\triceratops}{\textsc{triceratops}\,}
\newcommand{\tkmatrix}{\textsc{matrix}\,}
\newcommand{\rebound}{\textsc{rebound}\,}

\newcommand{\allesfitter}{\textsc{allesfitter}\,}
\newcommand{\juliet}{\textsc{juliet}\,}
\newcommand{\exoplanet}{\textsc{exoplanet}\,}
\newcommand{\sspe}{\textit{SSPE}\,}
\newcommand{\everest}{{\fontfamily{pcr}\selectfont everest}\,}
\newcommand{\kssf}{{\fontfamily{pcr}\selectfont \textit{\emph{K2}ssf}}\,}

\newcommand{\tapir}{\textsc{tapir}\,}

\newcommand{\lightkurve}{\textsc{lightkurve}\,}
\newcommand{\eleanor}{\texttt{eleanor}\,}
\newcommand{\sherlock}{{\fontfamily{pcr}\selectfont SHERLOCK}\,}
\newcommand{\watson}{{\fontfamily{pcr}\selectfont WATSON}\,}
\newcommand{\yaml}{{\fontfamily{pcr}\selectfont yaml}\,}
\newcommand{\csv}{{\fontfamily{pcr}\selectfont csv}\,}
\newcommand{\pdf}{{\fontfamily{pcr}\selectfont pdf}\,}

\newcommand{\candidates}{{\fontfamily{pcr}\selectfont candidates.log}\,}
\newcommand{\report}{{\fontfamily{pcr}\selectfont report.log}\,}
\newcommand{\fov}{{\fontfamily{pcr}\selectfont /fov}\,}
\newcommand{\sa}{{\fontfamily{pcr}\selectfont /sa}\,}
\newcommand{\rms}{{\fontfamily{pcr}\selectfont /rms}\,}
\newcommand{\detrends}{{\fontfamily{pcr}\selectfont /detrends}\,}
\newcommand{\fit}{{\fontfamily{pcr}\selectfont /fit}\,}

\title[The SHERLOCK pipeline]{The SHERLOCK pipeline: new exoplanet candidates in the WASP-16, HAT-P-27, HAT-P-26, and TOI-2411 systems}

\author[Dévora-Pajares et al.]{
Mart\'in D\'evora-Pajares,$^{1,2}$\thanks{E-mail: mdevorapajares@correo.ugr.es}
Francisco J. Pozuelos,$^{3}$\thanks{E-mail: pozuelos@iaa.es}
Antoine Thuillier,$^{5, 6}$
Mathilde Timmermans,$^{4}$
\newauthor
Val\'erie Van Grootel,$^{5}$
Victoria Bonidie,$^{7,8}$
Luis Cerde\~no Mota,$^{x}$
Juan C. Su\'arez,$^{1}$
\\
\\
$^{1}$Dpto. Física Teórica y del Cosmos. Universidad de Granada. 18071. Granada, Spain\\
$^{2}$Avature Machine Learning, Spain\\
$^{3}$Instituto de Astrof\'isica de Andaluc\'ia (IAA-CSIC), Glorieta de la Astronom\'ia s/n, 18008 Granada\\
$^{4}$Astrobiology Research Unit, Universit\'e de Li\`ege, All\'ee du 6 Ao\^ut 19C, B-4000 Li\`ege, Belgium\\
$^{5}$Space Sciences, Technologies and Astrophysics Research (STAR) Institute, Universit\'e de Li\`ege, All\'ee du 6 Ao\^ut 19C, B-4000 Li\`ege, Belgium \\
$^{6}$Institut d’Astronomie et d’Astrophysique, Université Libre de Bruxelles (ULB), CP 226 1050 Bruxelles, Belgium\\
$^{7}$Department of Physics and Astronomy, University of Pittsburgh, 3941 O’Hara Street, Pittsburgh, PA 15260, USA\\
$^{8}$Pittsburgh Particle Physics, Astrophysics, and Cosmology Center (PITT PACC), University of Pittsburgh, Pittsburgh, PA 15260, USA
}

\date{Accepted XXX. Received YYY; in original form ZZZ}

\pubyear{2024}

\begin{document}
\label{firstpage}
\pagerange{\pageref{firstpage}--\pageref{lastpage}}
\maketitle

\begin{abstract}
The launches of NASA \emph{Kepler} and \emph{TESS} missions have significantly enhanced the interest in the exoplanet field during the last 15 years, providing a vast amount of public data that is being exploited by the community thanks to the continuous development of new analysis tools. However, using these tools is not straightforward, and users must dive into different codes, input-output formats, and methodologies, hindering an efficient and robust exploration of the available data. We present the \sherlock pipeline, an end-to-end public software that allows the users to easily explore observations from space-based missions such as \emph{TESS} or \emph{Kepler} to recover known planets and candidates issued by the official pipelines and search for new planetary candidates that remained unnoticed. The pipeline incorporates all the steps to search for transit-like features, vet potential candidates, provide statistical validation, conduct a Bayesian fitting, and compute observational windows from ground-based observatories. Its performance is tested against a catalog of known and confirmed planets from the \emph{TESS} mission, trying to recover the official TESS Objects of Interest (TOIs), explore the existence of companions that have been missed, and release them as new planetary candidates. \sherlock demonstrated an excellent performance, recovering 98\% of the TOIs and confirmed planets in our test sample and finding new candidates. Specifically, we release four new planetary candidates around the systems WASP-16 (with P$\sim$10.46\,d and R$\sim$2.20\,$R_{\oplus}$), HAT-P-27 (with P$\sim$1.20\,d and R$\sim$4.33\,$R_{\oplus}$), HAT-P-26 (with P$\sim$6.59\,d and R$\sim$1.97\,$R_{\oplus}$), and TOI-2411 (with P$\sim$18.75\,d and R$\sim$2.88\,$R_{\oplus}$).

\end{abstract}




\section{Introduction}
The search for exoplanets has experienced one the most significant developments in modern astronomy, where space missions such as \emph{Kepler} \citep{kepler} and \emph{TESS} \citep{TESS} have emerged as beacons of discovery confirming thousands of new worlds and providing with thousands of planetary candidates waiting to be confirmed. In addition, these missions' public nature has produced hundreds of thousands of light curves, allowing the community to conduct alternative data exploration.  
To explore and exploit this vast amount of data, many software packages emerged; some designed for the analysis of light curves, including data loading and visualization tools such as \lightkurve \citep{lightkurve} and \eleanor \citep{eleanor}, 
some others to conduct data detrending using a variety of functions and algorithms to remove instrumental systematics that produce photometric variability such as \everest \citep{everest1,everest2}, \kssf \citep{k2ssf} and \wotan \citep{hippke2019}, others to conduct a transit search such as the Box Least Squares \citep[BLS;][]{bls}, the quasiperiodic automated transit search \citep[QATS;][]{qats}, and the Transit Least Squares \citep[TLS;][]{tls}, others to perform results vetting such as \vetter \citep{vetter} and \robovetter \citep{coughlin2016}, others dedicated to validate planetary candidates by determining whether a transit-like signal is statistically likely to be caused by a planet or a background eclipsing binary such as \vespa \citep{morton2012}, \pastis \citep{pastis} and \triceratops \citep{triceratops}, others to conduct a Bayesian fitting of the data to refine the planet parameters such as \allesfitter \citep{allesfitter}, \juliet \citep{juliet} and \exoplanet \citep{exoplanet2}, and others to compute observational windows such as \tapir \citep{tapir}, to name a few. 

While most of these packages are open-source, their different methodologies and input-output formats hinder their correct usage and implementation, relegating them to independent uses in most cases instead of an efficient workflow where all the procedures are connected. The lack of connection between these packages drives the community to rely mostly on the alerts issued by the official pipelines such as Science Processing Operations Center \citep[SPOC;][]{spoc} and the Quick Look Pipeline \citep[QLP;][]{qlp}, that is, the so-called \emph{Kepler} objects of interest (KOIs) and \emph{TESS} objects of interest (TOIs). However, many planets remain unnoticed due to the detection thresholds of these pipelines, the lack of data exploration, or the poor photometric quality. A proof of that is the number of alternatives catalogs providing hundreds of new planetary candidates using \emph{Kepler} data \citep[see, e.g.,][]{ofir2013,coughlin2016}, \emph{K2} data \citep[see, e.g.,][]{mayo2018,barros2016,kruse2019} and \emph{TESS} data \citep{nemesis,orion,diamante2020, diamante2023}, highlighting the importance of the parallel analyses of the data provided by the space-based missions by multiple teams. A brief review of some of these alternative catalogs can be found in \cite{melton2023}, where the authors presented their own catalog using \emph{TESS} data from cycle 1. Nonetheless, the pipelines producing these extra candidates are mostly non-public; only a few are open-source, such as \dave \citep{dave}, which is designed to search and vet planetary candidates using \emph{K2} data. Then, the community is restricted to solely focus on those alerts, impeding the freedom to concentrate on specific scientific cases designed by independent groups.
This paper presents the \sherlock (Searching for Hints of Exoplanets fRom Light curves Of spaCe-based seeKers) pipeline. This public Python package allows the users to conduct their own planetary searches, examining robustly the data from space-based missions. The tool was first introduced at its initial development stages with its core search functionality in \cite{pozuelos2020} and \cite{demory2020}, and has been used in a number of studies. The iterative development has led \sherlock to become an end-to-end pipeline where a special effort has been made to reduce user intervention and sorting all the steps in an efficient workflow. With only a few command lines, the users can explore the existing data, conduct the transit search, vet the most promising signals, compute a statistical validation, conduct a Bayesian fitting to refine planetary parameters and ephemerides, and trigger a follow-up camping using ground-based facilities. 
With the implementation of its six modules, \sherlock can be used in any imaginable exoplanetary survey that relies on space-based data. For example, \sherlock is being used in the SPECULOOS project \citep{gillon2018,sebastian2021}, which aims at finding transiting planets orbiting nearby ultracool dwarfs using a network of six 1-m ground-based telescopes located in different observatories such as ESO Paranal Observatory (Atacama Desert, Chile) \citep{burdanov2018,delrez2018}, Teide Observatory (Canary Islands, Spain) \citep{burdanov2022} and National Astronomical Observatory of Mexico (San Pedro M\'artir, Mexico) \citep{demory2020}. Due to the small sizes of these stars, transiting planets orbiting them are prime targets for atmospheric characterization using, for example, the \emph{James Webb Space Telescope} \citep{gillon2017,gillon2020}. In this context, \sherlock is exploring those stars in the SPECULOOS target list that have been observed by \emph{TESS}, searching for transit-like features that might be attributable to planets. While the SPECULOOS telescopes have better photometric precision than \emph{TESS} for these stars \citep[see, e.g.,][]{delrez2022}, a robust inspection of the \emph{TESS} data can help to speed up the process of planetary searching, saving telescope time.
In addition, \sherlock is the main pipeline of the FATE project \citep{vangrootel2021}, which aims at finding the first observational evidence of transiting planets orbiting hot subdwarf stars, which are direct post-RGB objects that have
lost most of their envelopes. In particular, the project objective is to shed light on understanding what happens to planetary systems once their host stars leave the main sequence and engulf close-in planets, presumably perturbing the dynamical architecture of the system. In this context, \sherlock systematically searches for transiting-like features in the \emph{Kepler}, \emph{K2}, and \emph{TESS} data \citep{thuillier2022}.

Hence, the idea behind \sherlock is not producing a vast amount of new planetary candidates as other existing pipelines do, but empowering independent researchers to carry on their own designed surveys.
The paper is organized in the following way:
Section~\ref{sec:sherlock} presents the key concepts and the workflow behind \sherlock; in Section~\ref{sec:recovery}, we use \sherlock to recover the planets of some known \emph{TESS} systems, evaluate the pipeline performance while also presenting the detection of four new candidates within those systems. In Section~\ref{surveys}, we present a new project in which \sherlock will be used to search for extra planets that remain unnoticed. Finally, Section~\ref{conclusions} presents our discussions, conclusions, and prospects. 

\section{The SHERLOCK pipeline}
\label{sec:sherlock}
\sherlock is an end-to-end pipeline that allows the users to explore the data from space-based missions to search for planetary candidates. It can be used to recover alerted candidates by the automatic pipelines such as Science Processing Operations Center \citep[SPOC;][]{spoc} and the Quick Look Pipeline \citep[QLP;][]{qlp}, the so-called \emph{Kepler} objects of interest (KOIs) and \emph{TESS} objects of interest (TOIs), and to search for candidates that remain unnoticed due to detection thresholds, lack of data exploration or poor photometric quality. To this end, \sherlock has seven different modules to 
(1) acquire and prepare the light curves from their repositories, (2) search for planetary candidates, (3) vet the interesting signals, (4) perform a statistical validation, (5) carry out planetary system stability tests, (6) model the signals to refine their ephemeris, and (7) compute the observational windows from ground-based observatories to trigger a follow-up campaign. To execute all these modules, the user only needs to fill in an initial \yaml file with some basic information such as the star ID, the cadence to be used, etc., and use sequentially one command line to pass from one step to the next. Alternatively, the user may provide \sherlock with the light curve in a \csv file, where the time, the normalized flux, and the flux error need to be given. In the following subsections, we describe in detail each of these steps.
The complete list of parameters that can be used in the \yaml file, along with a description of each one, can be found in GitHub's \sherlock repository\footnote{\url{https://github.com/franpoz/SHERLOCK/blob/master/sherlockpipe/properties.yaml}} keeping in mind that future implementations and releases might require new parameters; hence we suggest regularly checking the list.

\subsection{Light curve acquisition and preparation}
\label{sec:module1}

\sherlock makes use of the \lightkurve package \citep{lightkurve} to obtain the data of the space-based missions \emph{Kepler}, \emph{\emph{K2}} and \emph{\emph{TESS}}, from NASA Mikulski Archive for Space Telescope (MAST). In most cases, the user only needs to provide the star's KIC, EPIC, or TIC identification number and the cadence to be used, that is, 20, 120, 200, 600 or 1800~s for \emph{TESS} and 120 or 1800~s for \emph{Kepler}/\emph{K2}. In the case of the 600 and 1800~s for \emph{TESS}, that is, the full-frame images (FFIs), the data can also be retrieved via the \eleanor package \citep{eleanor}.
By default, the Pre-search Data Conditioning Simple APerture (PDC-SAP) is used, which has some level of processing with long-term trends removed using co-trending basis vectors. However, 
the PDC-SAP light curves are often affected by some other trends or systematics that need to be corrected before starting any planetary search. For example, some light curves present noisy regions with a high level of data dispersion, some others are affected by strong variability due to a fast rotation or stellar pulsations, and some others present short-term trends induced by the instrumentation. To efficiently deal with all these scenarios \sherlock accounts with different semi-automatic tools. 

\subsubsection{Masking noisy regions}
\label{rms}
\begin{figure}
	\includegraphics[width=\columnwidth]{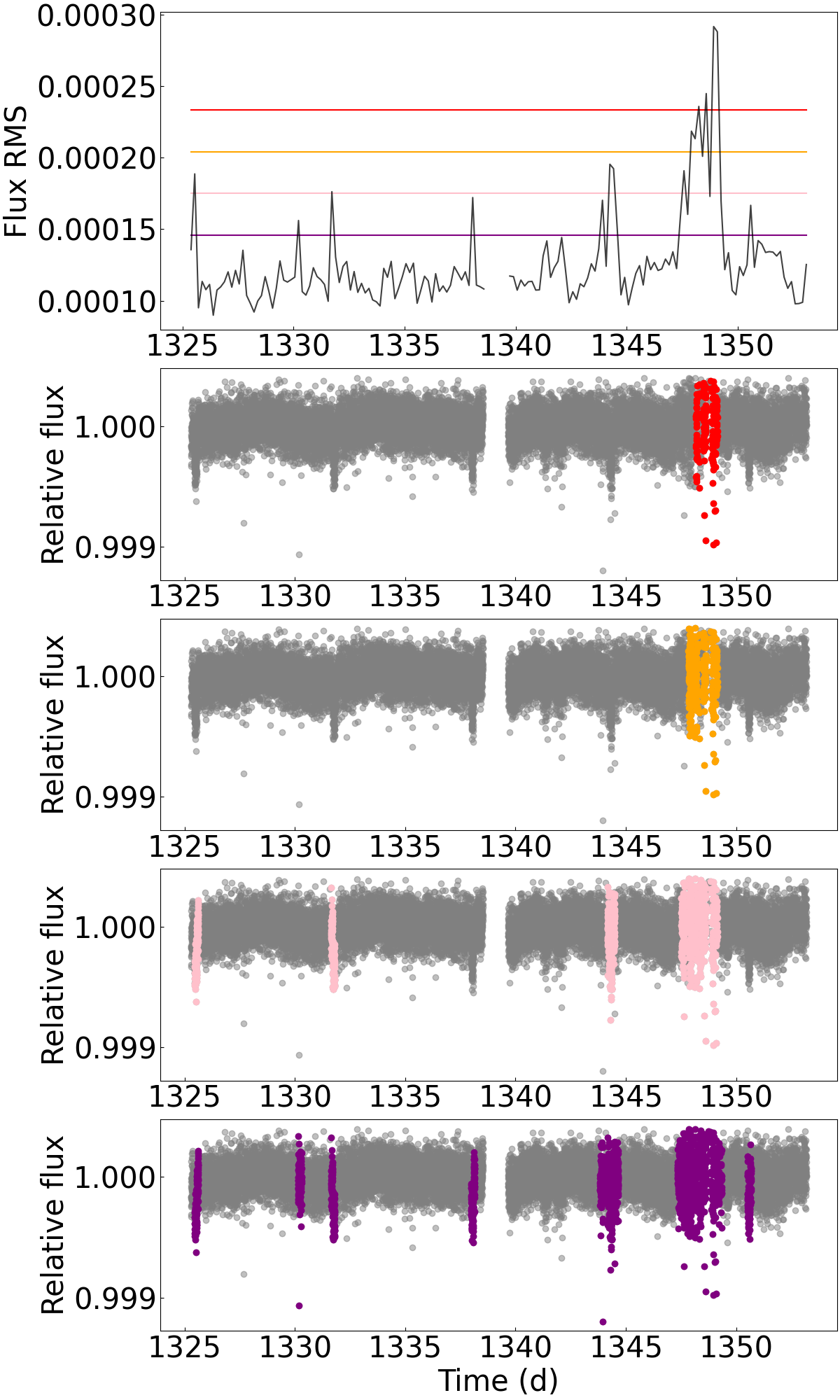}
    \caption{Output from \sherlock high-RMS masked regions for the \emph{TESS} light curve corresponding to TIC 261136679, observed during the Sector 1. The first panel shows the binned RMS values and the computed thresholds: red, the user-selected value (2 $\times$ RMS); purple, 1.75 $\times$ RMS; pink, 1.5 $\times$ RMS; orange, 1.25 $\times$ RMS. The bottom panels show the curve flux values masked by each of the shown thresholds in the top panel.}
    \label{fig:rms}
\end{figure}

The light curves often show regions with high data dispersion where the photometric quality is significantly reduced, strongly affecting the performance of the search algorithm. These noisy regions are sometimes related to the spacecraft's motion, light scattering produced by the Earth, background moving objects, or other unidentified sources. Hence, to clean the light curve of such nuisance data, \sherlock can automatically compute the root-mean-square (RMS) of the dataset by four-hour binning. Afterwards, all the light curve time gaps longer than three days are identified to split the data into segments. The mean RMS of each data segment is then computed, and each data point that falls into an RMS bin above 2$\times$ the mean RMS of its respective data segment is removed (see Fig.~\ref{fig:rms}). The threshold of 2 is applied by default, but the user can adjust it if needed in the \yaml\ file. As shown in Fig.~\ref{fig:rms}, \sherlock also plots three RMS threshold suggestions (1.25, 1.5, and 1.75 $\times$ the RMS) alongside the user selection to provide insight into the typical values used for this mask. In this figure, some potential transit signals from TOI-144 appear to be masked when low RMS threshold values are used, whereas a threshold of $1.75\times$ RMS seems to be optimal for masking the entire noisy region without affecting the transits. In cases where the automatic high-RMS-masking algorithm is not precise enough due to the complexity of the noisy features, the user can freely enter a set of time ranges to be masked.

\subsubsection{Stationary stellar pulsations extraction (\sspe)}
\label{sspe}
For quasi-stationary pulsations, we identified many cases (thanks to injection and recovery tests) where short-period transits might be confused with them, and then our targets would be removed using the automatic RMS-based noise removal algorithm. For such targets where these types of pulsations are identified, we developed a custom variation of the \cite{barcelo2015} steps to reduce their effect on the light curve:

\begin{enumerate}
\itemsep0em 
\item Clear transits masking: An initial batch of searches with \tls is done to mask transits with low residuals, which are typically appreciable in the light curve periodogram and might significantly alter the results of the next steps. That would happen because the sub-harmonics of the main transit period get narrower, and their shape in the periodogram might get close to the ones offered by the stationary pulsations that we would like to reduce.
\item Computation of highest-peak signal-to-noise (S/N): Once the highest peak is identified in the periodogram, its S/N is computed by dividing its power by the median of powers of its surrounding periods within a window of 4 hours. We used the median instead of the mean as it is typically a more robust statistical estimator of baseline values, which is what we try to extract here.
\item Model the stationary signal: In case the S/N is higher than a given threshold (defaults to 4), the signal is modeled with a non-linear least squares algorithm with an initial estimation of the amplitude and frequency through a fit of a cosine wave.
\item Check model validity: If the model shows an amplitude uncertainty lower than 20\% and a phase uncertainty lower than 0.2 radians, it will be extracted from the current light curve flux. If the resulting signal shows lower RMS, it will be accepted, and the resulting flux will be stored. We return to step 2 and apply the algorithm iteratively whenever a model is accepted.
\end{enumerate}

\begin{figure}
	\includegraphics[width=\columnwidth]{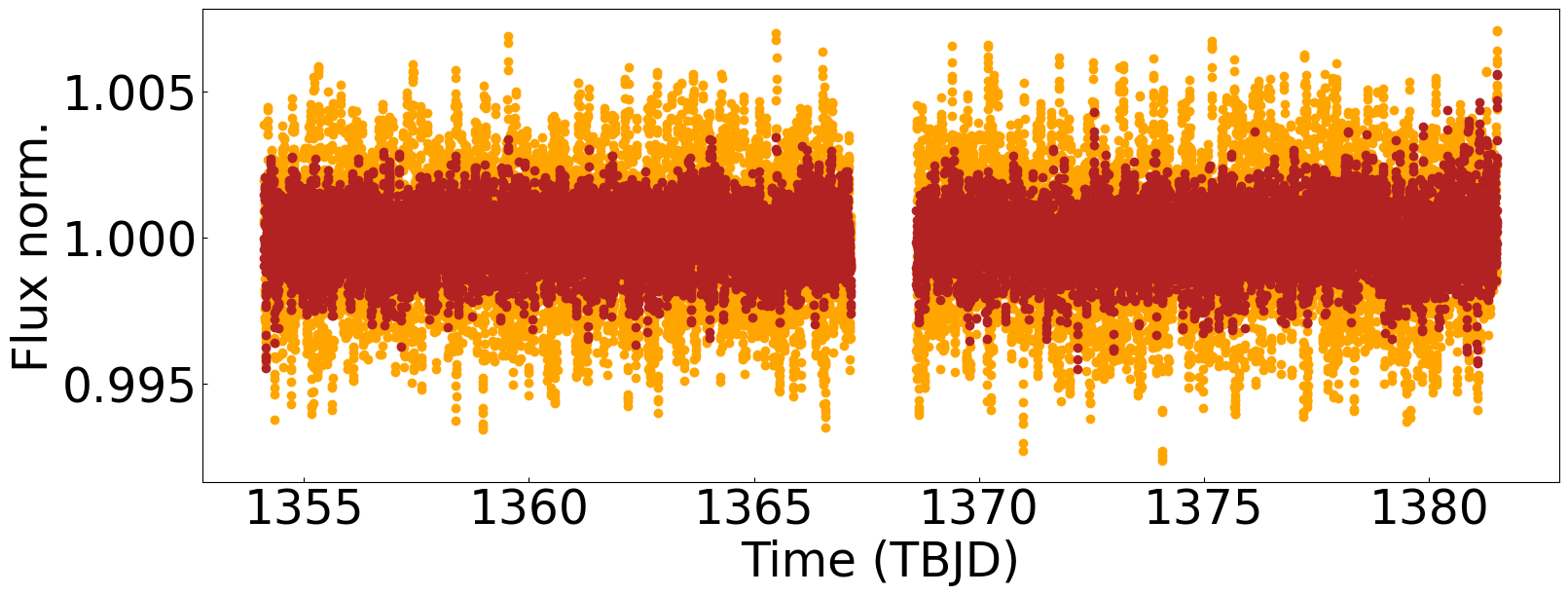}
    \caption{Results of applying the \sspe algorithm to the TIC 169285097 PDCSAP flux data obtained from TESS sector 2. The original light curve is depicted in orange, while the corrected light curve, following the removal of pulsations by our algorithm, is shown in red}
    \label{fig:sspe_flux_diff}
\end{figure}

To test the capabilities of the \sspe algorithm, we selected the TIC 169285097, cataloged as a hot subdwarf pulsating in p and g modes by \cite{sahoo2020}. These targets are complex hosts to search for small planets producing transits. Indeed, the strong RMS induced by the several pulsation modes would make such planets undetectable, increasing their residuals and reducing their S/N. Also, the typical transit detection algorithms often fail when searching for transits around these targets because they usually match sine-shaped pulsations to transit models, and their noise can be strongly correlated. As we are trying to diminish these effects, we tried our algorithm on the selected target to inspect the results that could be achieved on one of the available sectors from the TESS mission. We display the resulting flux in Fig.~\ref{fig:sspe_flux_diff}, which was obtained after subtracting 44 pulsation frequencies. The most powerful frequencies are shown in Fig.~\ref{fig:pulsation-modes}, where their corresponding models are overplotted to the phase-folded light curves. To properly investigate the capabilities of the algorithm in terms of transiting exoplanet signals recovery and to ensure that the procedure does not remove transit features, we tried our implementation on a public tool that we also developed in-house and published in a public repository under a free usage license: The MATRIX ToolKit\footnote{\url{https://github.com/PlanetHunters/tkmatrix}}. Built on top of the same code base as \sherlock, it contains the same algorithmic capabilities as the former and is conceived as an injection and recovery engine that includes a stage to generate a grid of periods, radii, and epochs to be injected in the original target's light curve. Those synthetic curves are later used in a recovery stage to search for the known injected transits. \tkmatrix will store a \csv file to keep track of which models were found and which were not, generating at the end a heat map showing the matches for each period, radius, and epoch from the grid.

\begin{figure}
	\includegraphics[width=\columnwidth]{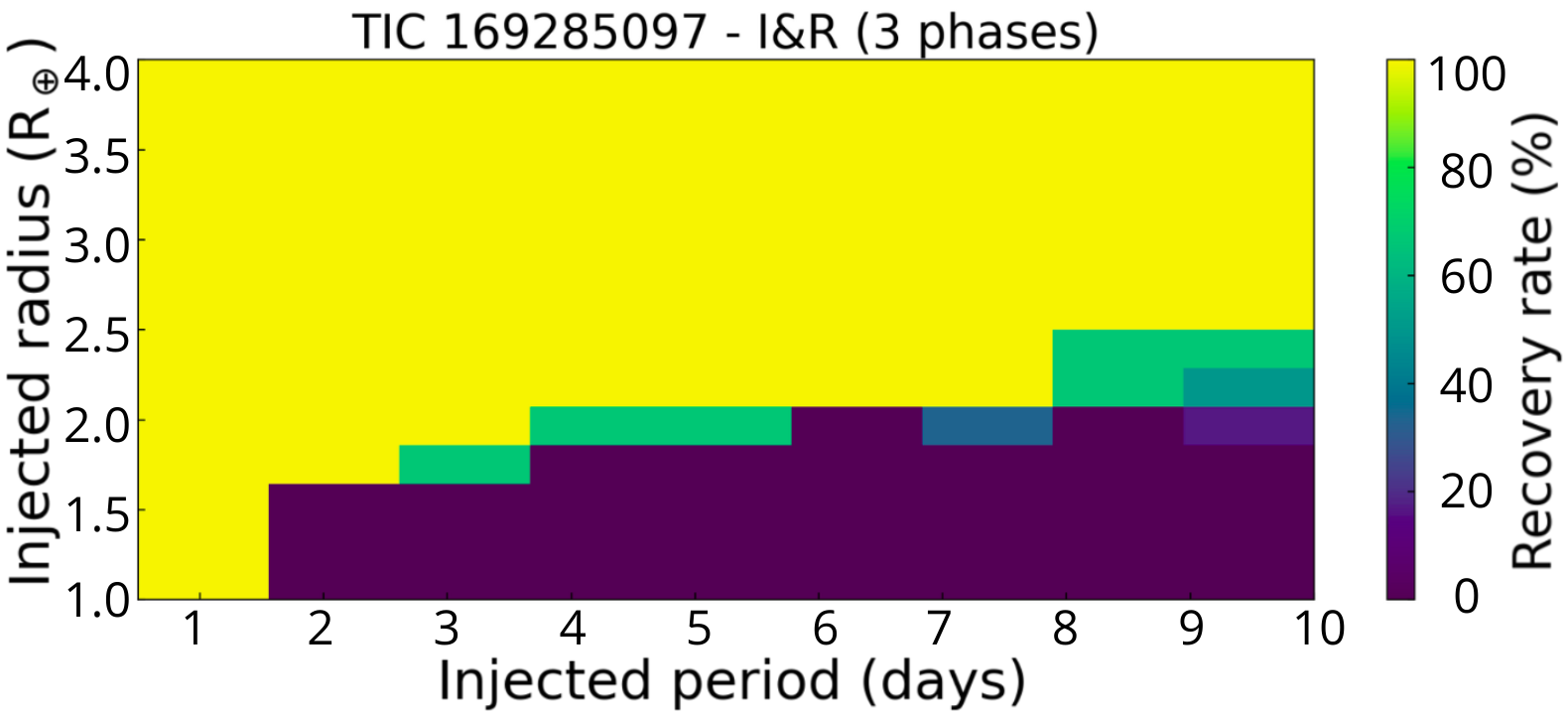}
	\includegraphics[width=\columnwidth]{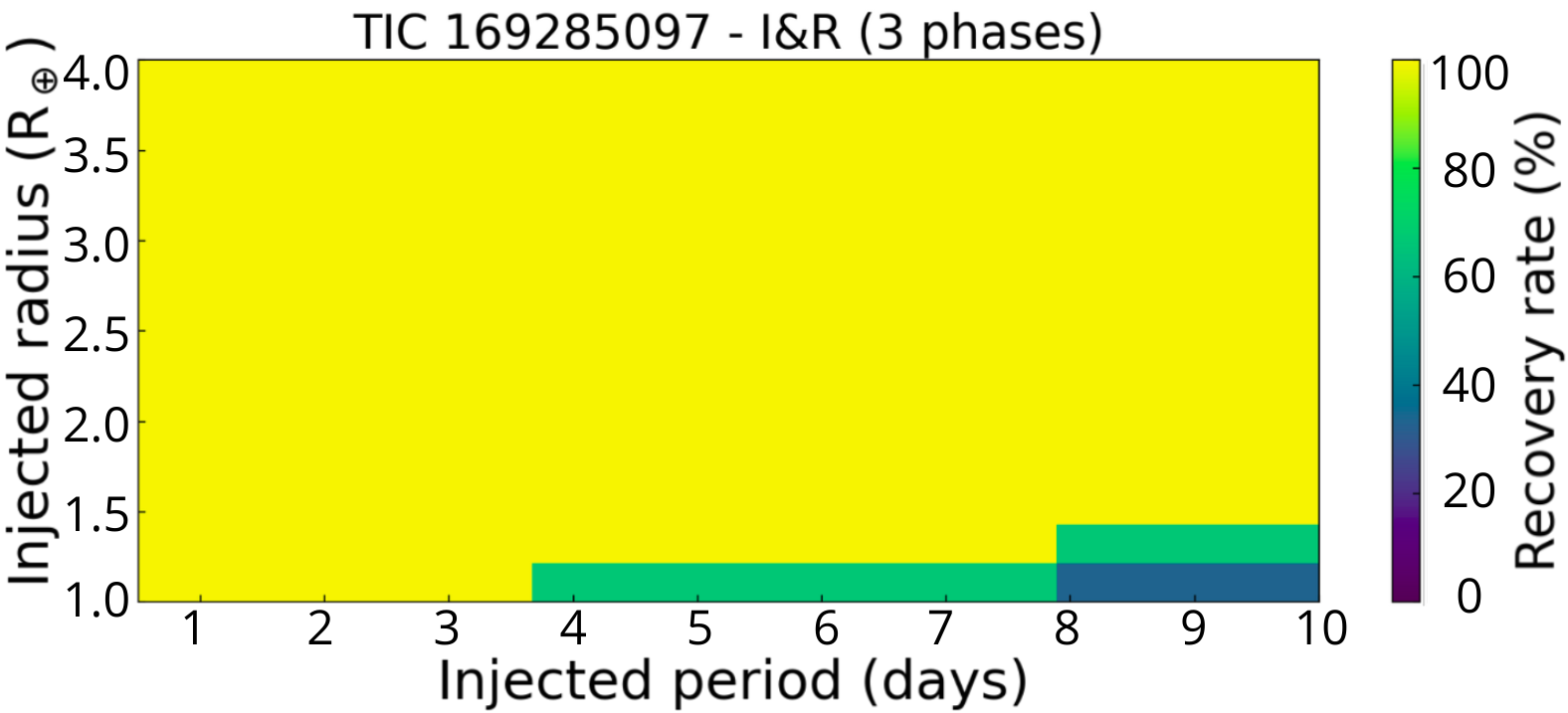}
    \caption{Injection and recovery results for the hybrid pulsator TIC 169285097 without and with the usage of \sspe. Above, the injection and recovery execution with a standard biweight detrend and search. Below, the same scenario execution including a previous \sspe run to reduce the quasi-stationary periodic pulsations effect on the light curve.}
    \label{fig:inj-rec-pulsations}
\end{figure}

To assess properly the improvement brought by our \sspe algorithm, we firstly tried the injection and recovery scenario for a simple search without using it. Later, we tried the same scenario with the inclusion of the algorithm. The results can be appreciated in the Fig.~\ref{fig:inj-rec-pulsations}, where we can see that the first algorithm could add for this target an increment in the detectability of $1$\,R$_{\oplus}$ for the best cases. Altogether this study case leads us to an outstanding improvement in our detection rates for our current transit surveys like the one started in \cite{vangrootel2021} and upcoming ones in the near future. 

An improved version of this iterative algorithm would include the entire set of already selected signals on each iteration to improve the model-fitted parameters as it is done with \felix in \cite{charpinet2020} and \cite{zong2016}. However, this is not included in the \sherlock \sspe procedure yet.

\subsubsection{Removing short trends}
\label{detrends}
\sherlock uses a multi-detrend approach to deal with short trends caused, for example, by instrumental drifts and intrinsic stellar variability. This strategy is implemented via the \wotan package \citep{hippke2019}. By default, \sherlock detrends the flux using the bi-weight method. This choice is based on its good speed-performance ratio compared with other methods, such as the Gaussian process, which is also available. The bi-weight method is a time-windowed slider, where short windows can efficiently remove short trends, but there is an associated risk of removing any actual transit signal. Hence, to prevent this issue, the pipeline explores, by default, ten cases with different window sizes. The minimum value of the window size is obtained by computing the transit duration (T$_{14}$) of a hypothetical Earth-size planet with an orbital period of 10~d. A minimum window size of 3$\times T_{14}$ is set to protect such a transit.
On the other hand, the maximum value explored is 20$\times T_{14}$, which is enough to remove the variability in most of the cases. However, if preferred, the user may set the number of detrends to be applied and the minimum and maximum values of the window size in the \yaml file. Hence, from the original PDC-SAP flux, \sherlock generates a new set of detrended light curves, each differing only in the window size used for detrending.

The transit search, described later in Section~\ref{search}, is then conducted over all the detrended light curves jointly with the original PDC-SAP light curve. If a signal is detected in only one (or very few) of the light curves, it is considered "detrend-dependent", suggesting it may be an artifact produced by the detrending algorithm. Conversely, if a signal is consistently found in multiple (or all) light curves, it is considered "detrend-independent", thus increasing its credibility. By comparing the S/N and Signal Detection Efficiency (SDE) of each detection across these light curves, we can determine the most appropriate window size for detrending. This method enhances the reliability of our planet detections by minimizing the risk of false positives due to the detrending process, ensuring that any promising signals found are not artifacts of a specific detrending configuration.

\subsubsection{Correcting strong variability}
\label{rot}
Many stars suffer significant changes in brightness, which might be caused, for example, by stellar spots or intrinsic stellar variability. In such cases, the PDC-SAP light curves show intense sinusoidal trends of high amplitudes. To deal with these scenarios, \sherlock 
computes the Lomb-Scargle periodogram (LSP) and extracts the value of the strongest peak using the \lightkurve package \citep{lightkurve}. By default, a factor of this value is used to remove the sinusoidal trend by fitting a sum of sines and cosines employing the \texttt{cosine} function provided by the \wotan package \citep{hippke2019}. Our calibration experiments showed that the most appropriate value for such a factor is 0.25. However, the user may adjust it to any other value if wanted in the \yaml file. There are some limitations in this procedure; for example, the presence of a hot Jupiter may induce a strong peak in the LSP that can be confused by a sinusoidal trend. Hence, a visual inspection is always desirable before applying this correction.

\subsection{Search for planetary candidates}
\label{search}
By default, \sherlock performs the transit search by using a modified version of the \tls package \citep[\tls ;][]{tls}. Since \tls uses an analytical transit model taking into account the stellar parameters, it offers a better S/N and SDE than classical algorithms based on \bls \citep[\bls ;][]{bls}. This stellar-model optimization allows the detection of shallow periodic transits, which are more challenging to find. However, if desired, the user may set the \bls algorithm in the \yaml file.

The procedure followed by \sherlock is to search for transits in all the detrended light curves jointly with the original one (PDC-SAP flux). Then, the results obtained for each individual light curve are stored; that is, the orbital period ($P$) and its error in days, mid-transit time for the first identified transit (T$_{0}$) in TBJD or BJD, the transit depth in ppt, the transit duration ($D$) in min, the number of transits, the S/N, the SDE, and the false-alarm probability (FAP). Once all the light curves are analyzed, the results are combined to select the best transiting signal. To this end, \sherlock provides first a \texttt{border-correction} algorithm. This algorithm computes a synthetic SDE value for each signal, penalizing the transits that are found in the border of the data set, which often corresponds to regions where there are more uncorrected systematics, as follow: 

\begin{gather*}
    SDE_{synthetic} = SDE \left(1 - \frac{ N_{tb}}{N_{t}}\right)
\end{gather*}
Where $N_{tb}$ is the number of transits close to borders and $N_{t}$ is the total number of transits. Secondly, the signal selection is performed by a \texttt{quorum} algorithm, where signals found in more light curves are favored in a voting process over those found on less light curves. 
Once a signal is selected as the best candidate, \sherlock masks it in the PDC-SAP light curve using its $P$, T$_{0}$ and 3$\times D$, and starts a new searching run. 
The user can set how many times this process is repeated in the \yaml file; however, our experience points out that results found beyond five runs are less reliable. This issue is coming due to the accumulated gaps in a given light curve after many mask-and-run iterations. Hence, by default, this process is repeated five times. In addition, \sherlock establishes minimum thresholds for the S/N and the SDE of a selected signal to pass from one run to the next. By default, both S/N and SDE are set to five, but the user may change these minimum thresholds in the \yaml file.

Besides the \texttt{quorum} algorithm, \sherlock has two other more simple alternative selection methods. These two methods are based on the selection of the signal with larger S/N and the larger S/N$_{synthetic}$, respectively. Moreover, \sherlock provides a flexible module to use any other user-defined selection algorithm.

Once the searching stage is finished, \sherlock produces as many folders as the number of runs executed, containing plots showing for each light curve (original PDC-SAP flux and detrended light curves) the best signal found. Moreover, \sherlock provides two log files: (1) \report, where the user can find all the information about the searching process, and (2) \candidates, where is listed a summary of the main signals found. Moreover, the user will have access to extra folders named \fov, \detrends, \rms and \sa. In \fov (from field-of-view) is stored the target pixel file (TPF) with the aperture used to extract the photometry over-plotted along with the nearby stars from Gaia DR2 catalog, obtained via the \tpfplotter package \citep{tpfplotter}, and the TPF with the light curves corresponding to each individual pixel using the \lightkurve package \citep{lightkurve}. In \detrends, the user can visually inspect all the detrends applied to the original PDC-SAP light curve as described in Section~\ref{detrends}. In \rms, a collection of plots shows the regions of the light curve that are masked due to their high RMS values, if any, as explained in Section~\ref{rms}. In case the \sspe module is activated (Section~\ref{sspe}), the extracted sinusoidal trends are plotted into the \sa directory. Finally, the LSP and the value of the strongest peak are displayed as described in Section~\ref{rot}. Combining these outputs, \sherlock gives the user the information needed to evaluate the results' reliability and the signals listed as promising.

\subsubsection{TLS metrics fine-tuning} 

The original \tls algorithm presented by \cite{tls} reported some inconsistencies with the SDE and S/N in some scenarios. \cite{tls} compute the transit duration by using a custom factor called `curve stretch`, supposedly taking into account the time gaps within the curve. Despite, this does not seem to be working properly and the error created in the transit duration propagates to the S/N calculation, which is based on this value. The S/N is especially hindered in curves with long time gaps, which are a typical case in the \emph{TESS} mission. That is why we decided to rewrite the transit duration calculation by only taking the best model fit provided by \tls, measuring the period fraction where it falls below 1. This fix to the algorithm seemed to repair the S/N computations and the package began to return very stable S/N and some compared outputs can be explored in Table~\ref{tab:tls_S/N}.

\begin{table}
    \centering
    \caption{S/N for three known transiting planets with long time gaps given by \sherlock searches with the standard \tls version and with our modified version.}
    \label{tab:tls_S/N}
    \begin{tabular}{lccr} 
        \hline  
        Planet & Sectors & S/N$_{TLS}$ & S/N$_{TLS-Modified}$\\
         \hline
        LHS 1140 b & 3,30 & 18.065 & 41.962\\
        TOI 178 d & 2,29 & 11.723 & 35.317\\
        TOI 736 c & 9,36 & 13.940 & 43.312\\
         \hline
    \end{tabular}
\end{table}

\begin{figure}
	\includegraphics[width=\columnwidth]{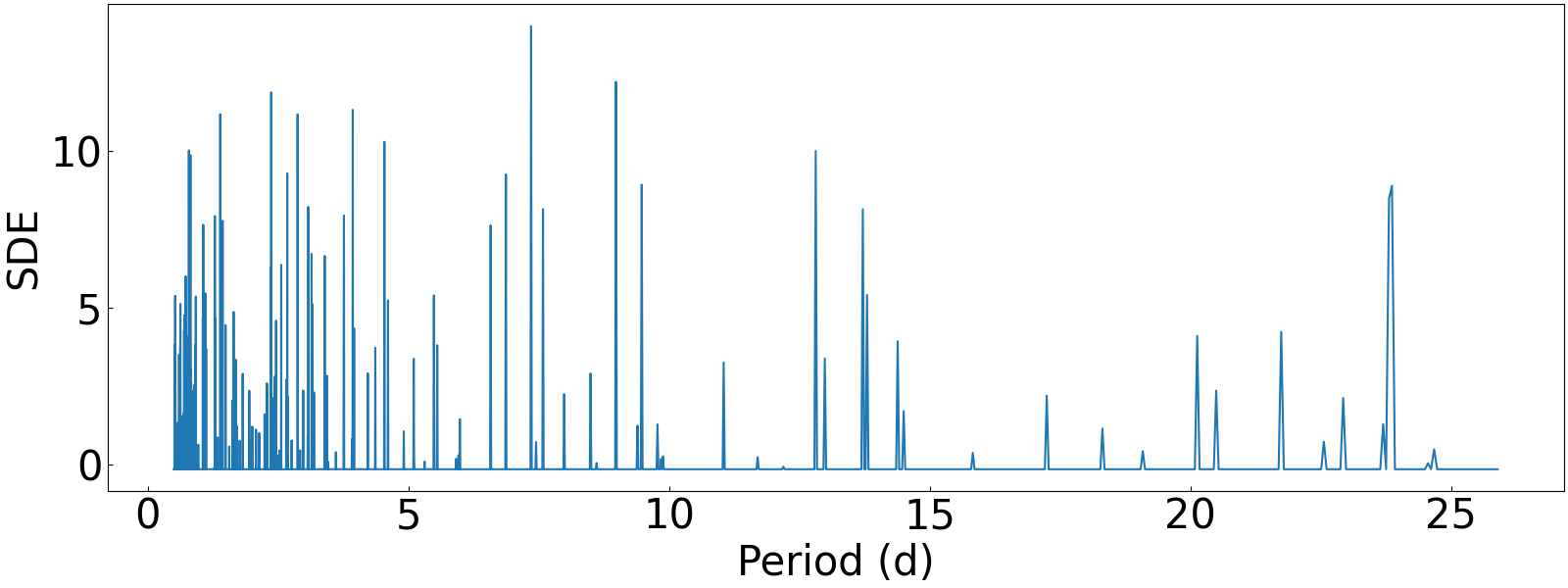}
	\includegraphics[width=\columnwidth]{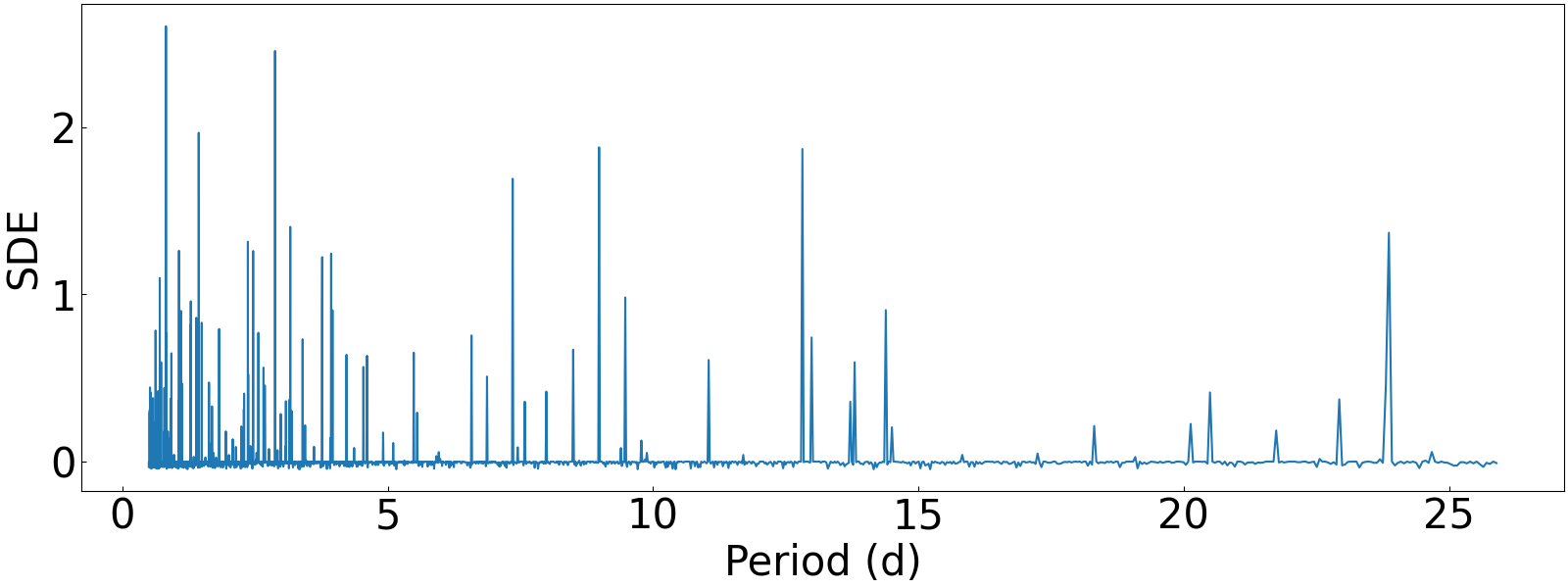}
    \caption{SDE periodogram for TIC 10837041 (TOI-2411) obtained during the third run, that is, the transits corresponding to TOI-2411\,b and another candidate found during the searching process (Sec.~\ref{sec:recovery}) have been removed from the light curve. Due to the short orbital period of TOI-2411 b, which is only 0.78 days, the resulting light curve used in the third run contains many gaps after removing the transits. These numerous gaps impact the computation of the SDE conducted by \tls, leading to overestimated values, as seen in the top panel. The bottom panel displays the proposed correction by \sherlock, where the SDE periodogram now displays more reliable values, indicating the very low detection significance of the found signals (see text for details).}
    \label{fig:sde}
\end{figure}

In addition, the SDE is a measurement that is expected to remain at scale among use cases and, hence, be comparable. We remind the reader that SDE is a metric indicating the statistical significance of a given period compared to the other periods tested in the transit search. In particular, an SDE value of $x$ for any given period $P$ means that the statistical significance of this period is $x\times\sigma$ compared to the mean significance of all other periods (we refer the reader to \cite{tls} and references therein for further details). Then, this metric allows us to evaluate and compare the prominence of detected signals across different periods. However, the computation of SDE by \tls implies the calculation of the residuals, which in some edge cases are set to a default baseline that can lead to an overestimation of the SDE. Specifically, when many residuals are set to the baseline value, the few significant spikes can cause abnormally high SDE values, for example, when a given light curve contains many gaps. To mitigate this, we rescale each final SDE value based on the proportion of local residuals that are above the baseline. This adjustment ensures that the SDE values more accurately reflect the true statistical significance of the detected signals. As an example, Fig.~\ref{fig:sde} shows the difference between an original SDE spectrum with many baseline values and its re-scaled counterpart. The last problem that we observed affecting the SDE is the difficulty of establishing fixed thresholds based on False Alarm Probabilities (FAP). The original \tls implementation calibrated a bijective relationship between SDE and FAP values. However, this calibration was carried out only for scenarios with timespans up to 28 days (the \emph{TESS} typical length for one sector data) and does not provide good FAP values for longer data contexts, where SDEs tend to be higher on average even for false positives.

\subsubsection{The Savitsky-Golay digital filter} 

\sherlock allows applying the Savitsky-Golay digital filter \citep{savgol} to the original PDC-SAP flux before any detrend and starting the transit search. This filter is implemented by using the \scipy package \citep{scipy}, which performs a data convolution by fitting sub-sets of adjacent data points with a low-degree polynomial using the linear least square method. Hence, the filter correlates nearby points in the data reducing their local white-noise without distorting the signal trend, leading to half the initial RMS value in the light curves. The Savitsky-Golay digital filter is handy when dealing with threshold-crossing events with low S/N. While the usage of this filter has the potential to find signals buried in the noise, it also has the associated risk of obtaining more false-positive detections that the user needs to vet carefully. In Table~\ref{tab:transit_params_comparison} we show the S/N and SDE values for TOI-270 \citep{gunther2019} and TOI-736 \citep{crossfield2019, peterson2023} using a pure-TLS search and applying first the Savitsky-Golay digital filter. As can be noticed, the S/N reported by TLS is improved by a factor of 2 and the SDE is sometimes improved and sometimes worsened without a noticeable impact in the search. The S/N and SDE reported by \sherlock are extracted directly from the \tls algorithm, calculated with a naive estimation that can not be compared with values from the SPOC pipeline. Therefore, these values are only used internally to compare the goodness of selected candidates and choose the clearest one. A recent application of this filter was used to recover the low S/N candidate TOI-4306.01 by \cite{delrez2022}, for which the pure-TLS search found the candidate with S/N below the typical detection threshold of S/N$\geq$5.

\begin{table}
    \centering
    \caption{Transit parameters for TOI-270 and TOI 736 known planets given by \sherlock searches without and with Savitzky-Golay filter.}
    \label{tab:transit_params_comparison}
    \begin{tabular}{lccccccc} 
        \hline  
        \multirow{2}{*}{Planet} & P & D & Depth & S/N & S/N & SDE & SDE\\
         & (d) & (min) & (ppt) & & (SG) & & (SG) \\
         \hline
        TOI-270 b & 3.36 & 75.3 & 0.8 & 14.3 & 31.6 & 61.1 & 40.4\\ 
        TOI-270 c & 5.66 & 94.7 & 2.9 & 49.1 & 112.4 & 39.8 & 46.6\\ 
        TOI-270 d & 11.38 & 115.9 & 2.0 & 28.1 & 70.2 & 44.7 & 44.0\\ 
        TOI-736 b & 0.95 & 35.8 & 3.8 & 8.9 & 17.4 & 18.5 & 18.7\\ 
        TOI-736 c & 4.99 & 59.1 & 14.4 & 17.7 & 41.2 & 24.4 & 25.0\\
        TOI-736 d & 2.75 & 49.2 & 2.15 & 5.192 & 12.2 & 6.6 & 6.6\\
         \hline
    \end{tabular}
\end{table}

\subsection{Semi-automatic vetting} 
\label{sec:vet}
As described in Section~\ref{search}, once the searching step is done, \sherlock provides several outputs that allow the user to perform a preliminary vetting. However, many other effects might be causing the detected signals that need careful inspection. 

The finished \emph{Kepler} mission \citep{borucki2003} provided more than 30,000 Threshold Crossing Events \citep[TCES;][]{thompson2018}, which are candidate signals that were found by the SPOC main search pipeline. However, different sources could generate these signals or even be instrumental artifacts induced into a target's light curve. To rule out these scenarios, the SPOC implemented the Data Validation Reports (DVR) \citep{twicken2018}, which are documents showing different metrics to qualify or discard an analyzed candidate. These scenarios are 1) transit shape model fit, 2) odd-even transits checks, 3) centroids shifts, 4) optical ghost effects, 4) transit source offsets, 5) rolling band contamination, 6) secondary depth significance, and some additional numerical tests. Based on these tests \sherlock implements a vetting procedure which are explained in the next sub-sections.

While these tests may offer a good vetting for promising signals, there are some limitations, mainly when dealing with shallow transits with very low S/N, where the results of the vetting might be inconclusive. \sherlock uses the vetting stage by importing the \watson package, which we also developed and published as an independent open-source tool\footnote{\url{ https://github.com/PlanetHunters/watson}}, offered in the official \python repositories\footnote{\url{https://pypi.org/project/dearwatson/}} with its own documentation\footnote{\url{https://dearwatson.readthedocs.io/en/latest/}}.

\subsubsection{Single transit inspection}
For each vetted candidate, \sherlock computes the light curve with the official pipeline aperture and a smaller one and displays them together for each single transit. The official pipeline aperture's centroids, motion, and background flux data are also displayed. In addition, the TPF is folded and plotted focused on each single transit. No metrics are computed from this data, but the visual inspection of these plots is very useful for finding differences or problematic features in single transits.

\subsubsection{Cadence-based curves signal reliability}
By using the period, epoch, and duration of a promising signal found in the transit search module, \sherlock computes the S/N for a box-shaped transit for each available cadence. If the transit is not significant enough (S/N$\geq$5) in any of the cadences or the S/N is substantially different from the nominal cadence, \sherlock will warn about it in the report.

\subsubsection{Odd-even and secondary events}
The transit-like signals might be originated by astrophysical sources that might not be planets. This is the case of Eclipsing Binaries (EB) or Nearby Eclipsing Binaries (NEB), which typically show secondary eclipses with different transit signatures to the primary one. \sherlock computes the S/N for odd and even transits depths and also for secondary eclipses. A red flag is reported when secondary eclipses are detected with S/N$\geq$3 or in case either odd or even transits show S/N$\leq$5.

\subsubsection{Transit source offset: images}
The difference imaging method proposed in \cite{bryson2017} was the selected approach by the SPOC to accept or reject signals based on the offset found on the computed transit source from pixel images. The method consists of computing the center of mass of the subtraction of the flux measured within the TPF for in-transit and out-of-transit per-pixel data. This way the most probable source of the transit events can be estimated. We have implemented a simplified version of this test in \sherlock. In addition, we have also included a second method to compute this metric by running a per-pixel \bls search with a fixed period on the folded TPFs. We then compute the so-called center of mass from each of the per-pixel \bls residuals and, finally, average the output of both methods to get the final transit source position.

\subsubsection{Transit source offset: centroids shift}
An alternative method to estimate scenarios where the transit signal might have originated outside the target is the centroids diagnosis, also presented by \cite{bryson2017}. In our case, we subtract the mission-computed motion values (polynomials relating the sky and the focal plane for each cadence) to the mission-computed centroid data (target position on the TPF given by the Pixel Response Function obtained from the out-of-transit control image \citep{bryson2013}). This method measures the deviation of the target position variations on the TPF against the focal plane alignment variations to the sky. If these values report an S/N > 3, \sherlock will be placing a red flag for the analyzed signals.

\subsubsection{Optical ghost diagnostic}
Sometimes the transit-like signals are caused by artificial sources. That is the case of optical ghosts, which can be created by light reflections from high-intensity sources between the optical system and the CCD as shown by \cite{caldwell2010}. To find false positives caused by optical ghosts, the \emph{Kepler} (DVR performs a test that measures the correlation between core (mean flux from all pixels within the target aperture) and halo (mean flux from all the pixels around the target aperture) fluxes and the transit signal. When the signal is affected by a ghost, the halo flux will be more correlated than the core one.

\sherlock implements a similar test. In our case, we compute and subtract the core and halo fluxes. In case the result still shows a transit signal with an S/N > 3, \sherlock will reject the candidate.

\subsection{System stability}
\label{sec:megno}

In some cases, several promising signals are selected during the transit search, but their orbital periods are very close to each other. In other cases, the selected signal is very close to a known confirmed planet. The system's architecture might be unstable in these scenarios, revealing that the candidates found are likely false positives.  

To shed light on these scenarios, \sherlock incorporates a module to assess the system dynamics stability. This analysis can be conducted at two points: during the initial vetting of a signal to assess whether there are possible system configurations for our candidates and as part of the validation scenario to get stability indicators for realistic or known configurations. That is, \sherlock can run a stability analysis in two modes. The first consists of giving user inputs, which create a linear grid of possible orbital parameters for each candidate. The second one loads the results from the Bayesian fit (see Section~\ref{sec:bayes}) for the selected signal, allowing the addition of some constraints to the most likely system configuration. 

To compute the system stability, \sherlock evaluates the fast chaos index Mean Exponential Growth factor for Nearby Orbits \citep[MEGNO;][]{cincottasimo1999,cincotta2000,cincotta2003}, which is widely used in this context \citep[see, e.g.,][]{jenkins2019,delrez2022,pozuelos2023}. \sherlock uses the N-body numerical package \rebound \citep{reboundMegno}, which provides a MEGNO implementation by employing the Wisdom-Holman WHfast code \citep{whfast}. The MEGNO mean value over time, denoted as $\langle Y(t) \rangle$, enhances stochastic fluctuations. This characteristic facilitates the differentiation of trajectories into quasi-periodic, that is to say, stable orbits, where $\langle Y(t\rightarrow \infty)\rangle$ approaches 2, and chaotic orbits, where $\langle Y(t\rightarrow \infty)\rangle$ tends towards infinity for an infinitive integration time or diverging from 2 for finite integration times.

This highly flexible module allows users to investigate as many parameters as desired to test the system's stability. Indeed, the user can explore the system's stability by varying the stellar mass, the planetary period, eccentricity, mass, argument of periastron, and inclination. Since we are dealing with transiting planets, not always the planetary masses are known; then, for these cases, \sherlock obtains the planetary mass relying on mass-radius relationships from \cite{chen2017}, which provides differentiation between small and large planets in a similar way than \cite{weiss2013} and \cite{bashi2017}. 
In all the cases, for each parameter \sherlock uses uniform distributions, where the user needs to provide the nominal value, the associate uncertainty, and the number of bins that want to explore within that range. The results are then stored in \csv files that the user can employ to construct stability maps. It is important to notice that while \sherlock offers the possibility of exploring a number of the parameters, not all of them equally affect the system stability. Indeed, as stated in \cite{cranmer2021}, the planetary semi-major axes (or the orbital periods), the masses, and the eccentricities are the parameters that most affect the stability of a given system. Out of these three parameters, usually are the masses and the eccentricities the most unconstrained and, hence, the ones used to explore system architecture \citep[see, e.g.,][]{demory2020,murgas2023}.          

To correctly sample the planetary positions over their orbits, the scenarios are integrated using a time-step of 5\% of the shortest orbital period in the system. Finally, running the integration for a time of $10^5$ to $10^6$ times the largest planetary orbital period in the system is recommended to start getting convergent and reliable results. In any case, this analysis does not intend to replace a more robust dynamical study of a given system but a preliminary view of the credibility of planetary candidates and the potential system architecture.   

\subsection{Statistical validation}
\label{sec:triceps}

\emph{Kepler}/\emph{K2} and \emph{TESS} found many candidates that, due to the faint nature of their host stars and/or the shallowness of their transits, can not be confirmed from 
ground-based observatories as true planets. To mitigate this situation were developed the first statistical-validation tools during the \emph{Kepler}-era
\citep[see, e.g.,][]{torres2011,morton2012,morton2016}, which allow to calculate the false positive probability of a given detection to validate it statistically.

This procedure is nowadays widespread to support the planetary interpretation of a given signal even when the candidate 
can be reached from ground-based observatories \citep[see, e.g.,][]{kostov2019,quinn2019}. \sherlock allows the user to perform a statistical validation automatically connecting the results with the \triceratops\footnote{\url{https://github.com/stevengiacalone/triceratops}} package \citep{triceratops,giacalone2022}. Originally, \triceratops was designed to vet and validate \emph{TESS} candidates only; however, we have contributed to developing this open-source package to manage also \emph{Kepler}/\emph{K2} candidates. In short, \triceratops computes the probabilities of a wide range of transiting-producing astrophysical scenarios such as a transiting planet and eclipsing binary (in many different configurations) using the primary transit of the candidate, the prior knowledge of the host and nearby stars, and the current understanding of planet occurrence and stellar multiplicity.    
Hence, \triceratops computes the location of a given candidate in the false-positive probability (FPP) -- nearby false-positive probability (NFPP) plane, where the candidate will be
classified as a validated planet if FPP$<$0.015 and NFPP$<10^{3}$, as likely planet if FPP$<0.5$ and NFPP$<10^{-3}$, and as likely false positive if NFPP$>10^{-1}$. Due to the 
probabilistic nature of the procedure to calculate the values of FPP and NFPP by \triceratops, \sherlock performs five computations and then averages the results to provide the final values. In addition, \sherlock uses the results from \cite{lissauer2012}, which estimated the fraction of planet candidates that are false positives when in systems with multiple planet candidates
based on \emph{Kepler} data. Then, this prior probability is multiplied by the FPP and NFPP previously computed \citep[see, e.g.,][]{demory2020}.

\subsection{Bayesian model to refine the ephemeris}
\label{sec:bayes}
Once a candidate has been vetted and validated, it can be desirable to trigger a follow-up campaign to firmly confirm the planet in the target star using ground-based observatories. 
However, the results from the searching stage via \tls are not optimal for scheduling an observation due to the large uncertainty associated with the found orbital period ($\Delta P$), whose value directly depends on the period grid density. Indeed, the transit time ($T$) and its uncertainty ($\Delta T$) of future observational windows are given by \citep[see, e.g.,][]{mallonn2019}:  

\begin{gather*}
   T= T_{0}+n\cdot P \\
   \Delta T=\sqrt{\Delta T_{0}^{2} + (n\cdot\Delta P)^{2}}  
\end{gather*}

With T$_{0}$ and $P$ previously defined in Section~\ref{search}, and $n$ the number of cycles since T$_{0}$. Hence, error propagation is strongly dominated by the term $n\cdot\Delta P$, 
which may produce $\Delta T$ of the order of several hours if the detection is performed in \emph{Kepler}/\emph{K2} data or early \emph{TESS} sectors. This situation might hinder any ground-based confirmation.

In this context, \sherlock allows the user to refine the ephemeris by performing a Bayesian fit using the \allesfitter\footnote{\url{https://github.com/MNGuenther/allesfitter}} package \citep{allesfitter-code}. \allesfitter is an open-source code that performs a Bayesian fit to infer the star and exoplanet properties given a set of photometric and/or radial velocity data. Then, \sherlock automatically injects the results from the searching stage and the stellar parameters in \allesfitter as priors and implements two steps fitting: 1) The candidate signals are masked, and the out-of-transit correlated noise estimation is performed using a Gaussian Process (GP) with a 3/2 Matérn kernel with a MCMC fit on the PDCSAP-flux light curve, which can describe smooth long-term trends and short-term stochastic variations by its two hyper-parameters: the amplitude scale ($\sigma$) and length scale ($\rho$). 2) The candidate signals priors are fit taking into account the previously derived GP parameters using normal priors, modeling the radius ratio of the planet to host, $R_{p}/R_{\star}$, the sum of the stellar and companion radii divided by the semimajor axis, $(R_{\star}+R_{p})/a_{b}$, the cosine of the orbital inclination, $\cos i_{p}$, the epoch/transit midtime in days, $T_{0}$, the orbital period of the planet in days, $P_{p}$, the transformed limb-darkening coefficients following a quadratic law, $q_{1}$--$q_{2}$, and the natural logarithm of the white noise, log$\sigma_{w}$.        
By default, the Bayesian model used by \sherlock is the Nested Sampling, but the user might switch to MCMC if desired in the \yaml file. All the posterior distributions and the derived parameters are stored in a new folder named \fit.
At the end of this stage, the user obtains a robust inference of the planetary parameters. In particular, the $\Delta P$ is reduced considerably, and the new values can be used to schedule ground-based observations with much more accurate predictions.

\subsection{Ground-based observational windows}

Once the planetary properties are optimal from Section~\ref{sec:bayes}, the user may safely plan a follow-up campaign from ground-based observatories to confirm the planetary nature of a given candidate. This step is particularly critical for results obtained using \emph{TESS} data. Indeed, the pixel scale of \emph{TESS}, of 21", is remarkably larger than \emph{Kepler}'s, of 4", moreover the \emph{TESS}' point-spread function (PSF) might be as large as 1'. Hence, due to dilution effects, the pixel scale, and the large PSF, the probability of contamination by nearby eclipsing binaries, whose deep eclipses might mimic shallow transits detected in the target star, is not null and, ideally, has to be checked from ground-based facilities. This ground-based confirmation is the primary goal of the seeing-limited photometry working group of \emph{TESS} \citep[see, e.g.,][]{kostov2019}. To schedule these observations, the community typically uses the \emph{TESS} Transit Finder tool, which is a customized version of the \tapir package \citep{tapir}.    

In this context, \sherlock automatically loads the Bayesian fit results and a list of observatories provided by the user from where the transit windows will be computed. The observational windows from each observatory will consider some user-defined constraints, such as the minimum altitude of the object, distance from the moon in relationship with its illumination phase, and the minimum transit coverage. Then, \sherlock displays the observational windows in \csv and \pdf files, including the ingress, mid-transit, and egress times in UT for each event and plots with the airmass and altitudes of the host star during the entire event. This module was built by using the \astropy package \citep{astropy1, astropy2} and was tested against the \emph{TESS} Transit Finder tool to verify its good performance.

\section{SHERLOCK performance}
\label{sec:recovery}

Previously, we have outlined several projects leveraging the \sherlock pipeline, such as SPECULOOS and FATE. Moreover, many other studies employed this pipeline to retrieve alerts from the TESS mission and known planets, as well as to identify new planetary candidates \citep[see e.g.,][]{wells2021,schanche2022,murgas2023,dransfield2023,dreizler2024,gillon2024}. In this section, our focus shifts to demonstrating the performance of the \sherlock pipeline through a specifically designed experiment to provide a level of reliability.

In particular, we built a catalog of 100 stars that fulfill the following conditions in the ExoFOP list\footnote{\url{https://exofop.ipac.caltech.edu/tess/view_toi.php}; retrieved in October 2023}: (1) the targets should, at least, contain one previously known planet or a TOI confirmed planet, (2) one of these planets should have orbital periods shorter than ten days, (3) the maximum number of sectors should be two, and (4) the produced list with previous conditions is sorted by descending brightness and limited to the first 100 stars.
Filters (1), (2), and (3) are applied to reduce the computational cost of this experiment. Appendix~\ref{app:recall_catalogue} lists the catalog obtained following these criteria providing the TIC-IDs.
The main goal of this test is to run \sherlock over this catalog and record the successful recovery and non-recovery of planets and TOIs. Hence, to easily illustrate the performance, we introduce the recall parameter defined as $TP / (TP + FN)$, where $TP$ are the true positives, that is, the number of correct planets or TOIs found, and $ FN$ are the false negatives, that is, the number of known planets or TOIs that we do not recover. Then, the performance of the code is optimal when this parameter approaches 1, indicating that \sherlock successfully identifies the expected planets without omissions.    

From the 100 stars composing our catalog, there were 104 TOIs that might correspond to previously confirmed planets, that is, known planets for which TESS also provided an official TOI, or new planetary candidate alerts released by TESS that the scientific community has confirmed as planets. The number of multi-alert stars was very reduced because of the filters used to build the target list. Indeed, our filters favored targets with large planets in short orbital periods, such as hot Jupiters (periods up to 10 days and radius from 6\,$R_{\oplus}$). Population studies suggested that nearby planetary companions rarely accompany these planets. However, thanks to the extreme photometric precision of space-based missions and dedicated radial velocity surveys, nearby planetary companions to hot Jupiters have emerged in recent years \citep{wu2023}. These recent discoveries suggest that the formation of hot Jupiters via quiescent mechanisms is more efficient than previously thought, when the main formation channel invoked was the dynamical high-eccentricity migration \citep{dawson2018}. Hence, in this context, our target list not only serves to compute the recall factor and establish the goodness of the \sherlock pipeline but also offers an interesting scientific case searching for low S/N signals that might hint at the existence of planetary companions to these hot Jupiters in our list, supporting this new emergence of planets.

Once our target list was well established, we ran \sherlock over it. In general, after the execution of the first module, that is, the exploration of the available data for a given candidate (see Section~\ref{sec:module1}), we triggered a standard search with no initial fine-tuning. However, we had to refine the search parameters for some targets due to several factors, such as poor photometric quality and lack of data from the official pipelines. We provide a brief explanation for these particular cases: 

\begin{itemize}
    \item TIC 146520535 and TIC 464646604: In our multi-detrend procedure, the window size ranges from 0.2 to 1.2 days by default, which we found optimal for most cases. However, due to the strong short-term variability of these light curves, we had to adapt the minimum and maximum window sizes to lower values from 0.25 to 0.75 days.
    \item TIC 189380158, TIC 281408474, and TIC 33595516: The data for these targets were not available in MAST, impeding us from gathering them using \lightkurve, which is the default route to get the data. Hence, we switch to \eleanor long-cadence data by specifying the author in the \yaml file.
\end{itemize}

\sherlock recovered 102 out of 104 transiting planets in the catalog, making a count of two false negatives. These two were Kepler-89\,e (TOI 4581.02), which was present in the TESS data as a single transit and hence, \sherlock mismatched it with a different period, and K2-285\,c (TOI 4538.01), which was entirely missed in favor of a different non-promising signal. With this data, the final recall made up a value of 0.98 for our catalog. However, \sherlock was also able to recover two other known planets that were not alerted with a TOI: the TOI-942\,c and Kepler-89\,c. For the TOI-942 system, TESS alerted one candidate with an orbital period of 4.3\,d, (TOI-942\,b), and the follow-up of the system conducted by \cite{carleo2021} found an outermost transiting planet with an orbital period of 10.16\,d (TOI-942\,c), which corresponds to our detection. In the case of the  Kepler-89 system \citep{marcy2014}, TESS alerted one candidate that corresponded to Kepler-89\,d with an orbital period of 22.34\,d, and a single transit that belonged to Kepler-89\,e. In our experiment, \sherlock also found a planetary candidate with an orbital period of 10.42\,d, which corresponded to Kepler-89\,c. 

\begin{figure*}
	\includegraphics[width=\textwidth]{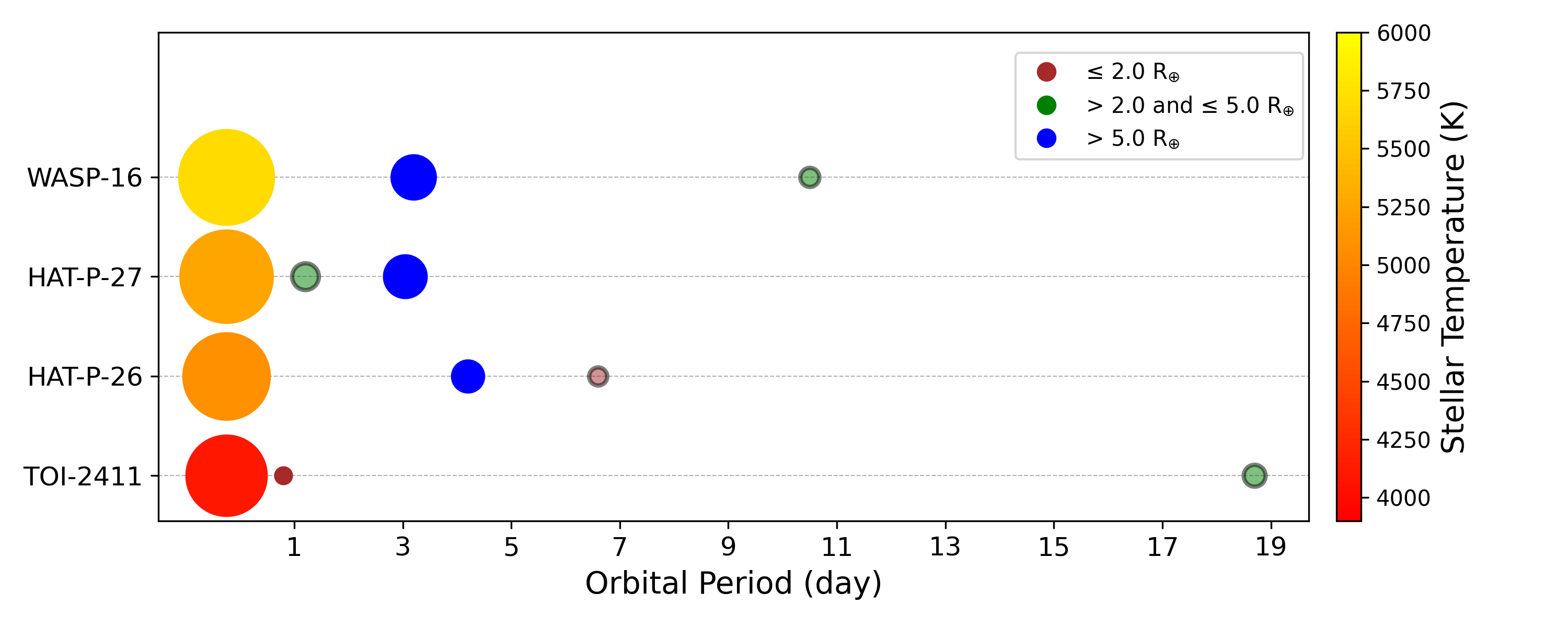}
    \caption{The planetary architectures of the systems WASP-16, HAT-P-27, HAT-P-26, and TOI-2411 are shown from top to bottom. The sizes of the elements in this figure scale with the sizes of the represented stars and planets. The sizes of the stars are scaled relative to each other, and the planets are scaled relative to each other, but the relative scales between stars and planets are not accurate, with planets oversized for clearer visualization. Known planets in the systems are displayed in solid colors, while the candidates found in this study are highlighted with a gray halo and transparency, whose properties are listed in Table~\ref{tab:fn_candidates}.}
    \label{fig:architectures}
\end{figure*}

On top of this, \sherlock spotted four interesting signals that could be attributable to planets and for which there is not any previous information. These new candidates were found in the systems WASP-16, HAT-P-27, HAT-P-26, and TOI-2411. We followed all the steps to verify their planetary nature by using the modules available in \sherlock as described in Section~\ref{sec:sherlock}. In the following subsections, we describe these new candidates and the confirmation steps that we conducted to claim them as planetary candidates with high probabilities of being genuine planets. Table~\ref{tab:fn_candidates} shows the summary of these new candidates, and the planetary systems architectures are presented in Fig.~\ref{fig:architectures}. We release these candidates as Community TESS Objects of Interest (CTOI), and will be listed in the ExoFOP CTOIs list\footnote{\url{https://exofop.ipac.caltech.edu/tess/view_ctoi.php}}.  

Hence, despite the small size of our sample, these results allow us to confirm the high performance of \sherlock as a planetary search engine that can be used to explore space-based data efficiently, not only to recover known planets but also to search for new ones that remained unnoticed by the community.

\begin{table*}
    \centering
    \caption{New planetary candidates found during the performance survey}
    \label{tab:fn_candidates}
    \renewcommand{\arraystretch}{1.5} 
    \begin{tabular}{lccccccr} 
        \hline  
        System & CTOI-ID  & Period (d) & T$_0$ - 2457000 (BJD$_{TDB}$) & Duration (h) & Depth (ppt) & Radius ($R_\oplus$) & T$_{eq}$ (K)\\
         \hline
        WASP-16 & 46096489.02 & $10.457_{-0.028}^{+0.018}$ & $1618.725_{-0.014}^{+0.012}$ & $3.67_{-0.45}^{+0.32}$ & $0.54\pm0.1$ & $2.20\pm0.23$ & $810\pm36$\\
        HAT-P-27 & 461239485.02 & $1.1994\pm0.0002$ & $2705.510_{-0.001}^{+0.002}$ & $1.27_{-0.10}^{+0.11}$ & $2.13\pm0.20$ & $4.33\pm0.44$ & $1426\pm87$ \\
        HAT-P-26 & 420779000.02 & $6.594_{-0.007}^{+0.009}$ & $2683.160_{-0.010}^{+0.014}$ & $2.71_{-0.41}^{+0.19}$ & $0.59\pm0.11$ & $1.97\pm0.20$ & $785\pm15$ \\
        TOI-2411 & 10837041.02 & $18.750\pm0.004$ & $2141.315_{-0.003}^{+0.004}$ & $3.56_{-0.08}^{+0.10}$ & $1.77\pm0.12$ & $2.88\pm0.15$ & $430\pm15$\\
         \hline
       
    \end{tabular}
\end{table*}

\subsection{WASP-16}
\label{sec:wasp16}
WASP-16 is a Sun-like star with $T_{\textrm{eff}}=5700\pm150\,K$ , $R_{\star}=0.946\pm0.054\,R_{\odot}$ and $M_{\star}=1.022\pm0.101 M_{\odot}$ observed by the WASP project \citep{wasp2006}, in particular using WASP-South, between 2006 and 2008. These observations hinted at the presence of a giant planet that was later confirmed using spectroscopic observations with the CORALIE spectrograph in 2008 and 2009. The results confirmed the planetary nature of this signal, yielding $R=1.008\,R_{Jup}$, $M=0.855\,M_{Jup}$ and with an orbital period of  3.12\,d. This planet is the hot Jupiter WASP-16\,b \citep{lister2009}.      

WASP-16 was observed by the TESS mission in Sector 11 between April and May of 2019, using the 120 and 1800\,sec cadences (TIC-46096489). The search performed with \sherlock was conducted using the standard parameters, and we recovered in the first run the signal corresponding to WASP-16\,b, with S/N=97.9 and SDE=18.6 in the 11 light curves  (see Section~\ref{detrends}).
During the second run, \sherlock did not spot any promising signal, but events that happened close to the borders of the light curves, which clearly hinted at a systematic origin. During the third run, we found a weak signal with S/N=4.7 and SDE=13 with a period of 10.44\,d. While this signal is just below our default detection threshold of S/N$\geq$5, it was found in all the 11 light curves processed, which reveals a non-detrend dependence. Hence, we vetted the signal and excluded the possibility of it being produced by any systematics (see Section~\ref{sec:vet}). Then, we conducted the statistical validation (see Section~\ref{sec:triceps}), and we got FPP=0.26 and NFPP=0.0036 values, slightly above the likely planet area defined by \cite{triceratops}. Subsequently, we conducted the Bayesian fitting (see Section~\ref{sec:bayes}) and compared the Bayes Factor ($\Delta \ln Z$) of two models: 1) only with the WASP-16\,b planet, that is, the null hypothesis, and 2) including the second potential planet found by \sherlock, that is, a more complex scenario with two planets. Then, we got $\Delta \ln Z = \Delta \ln Z_{2planet} - \Delta \ln Z_{1planet}\sim6.7$, what favours the 2-planet scenario \citep{kass1995}. Considering the global results of these analyses, we concluded that the signal found with a 10.44\,d is a strong planetary candidate named     
 CTOI-46096489.02, which would correspond to a mini-Neptune of $\sim$2.2\,R$_{\oplus}$. Its main parameters from the fitting can be consulted in Table~\ref{tab:fn_candidates}. 
However, it is important to notice that the Bayes Factor is suggestive but not decisive; hence, extra observations using high-precision photometry would be needed to firmly confirm the planetary nature of the CTOI-46096489.02 candidate. Due to the shallowness of its transit of $\sim$0.54\,ppt, ground-based instrumentation is inadequate for detecting such a subtle eclipse. Indeed, the limited sensitivity and atmospheric interference associated with terrestrial telescopes prevent accurate measurement of these weak signals. Therefore, to obtain precise and reliable data that allow us to confirm the planet, it is essential to utilize space-based instrumentation, such as CHEOPS \citep{cheops}, which can bypass atmospheric distortions and provide the necessary sensitivity to observe these shallow transits effectively \citep[see, e.g.,][]{delrez2021}. Unfortunately, since the last observational data available comes from TESS Sector 11, the current uncertainty in the transit time, $\Delta T$ (see Section~\ref{sec:bayes}), is $\sim$100\,h, which makes any time-critical observation impracticable. Another TESS-like campaign would be needed to mitigate this lack of accuracy in the current transit times. According to the TESS-point web tool\footnote{\url{https://heasarc.gsfc.nasa.gov/wsgi-scripts/TESS/TESS-point_Web_Tool/TESS-point_Web_Tool/wtv_v2.0.py/}}, WASP-16 will be re-observed in April 2025, what will offer the opportunity to recover this candidate and trigger an accurate follow-up campaign that will allow us to confirm or refute its planetary nature.   

The two-planet model is displayed in Fig.~\ref{fig:all_fit}, the posterior distribution is provided in Appendix~\ref{app:posterior_fitting}, Table~\ref{app:fits}, and the convergence of the model is visualized in Fig.~\ref{app:wasp16_corner}.

\subsection{HAT-P-27}
\label{sec:hatp27}

HAT-P-27 is a G8-dwarf star with a $T_{eff}=5250\pm100\,K$, $R_{\star}=0.90\pm0.05\,R_{\odot}$ and $M_{\star}=0.94\pm0.04\,M_{\odot}$ monitored on a nightly basis by The Hungarian-made Automated Telescope Network \citep[HATNet;][]{bakos2011} from January to August 2009. These photometric observations were combined with nine
spectroscopic measurements using the HIRES instrument on the Keck I Telescope from December 2009 to June 2010. The results were presented in \cite{beky2011}, reporting the discovery of HAT-P-27\,b, a hot Jupiter with a size of 1.038\,R$_{Jup}$, a mass of 0.66\,M$_{Jup}$ and an orbital period of 3.04\,d.    

HAT-P-27 was observed by the TESS mission in Sector 51 in May 2022, using the 120 and 600\,sec cadences (TIC-461239485). We used the 120\,s data in our search with \sherlock, allowing us to recover in all the light curvers a transiting signal corresponding to the HAT-P-27\,b planet with S/N=37.3 and SDE=12.6. During the second run, \sherlock spotted a transit-like feature with a period of 1.20\,d, with S/N=10.3 and SDE=29.8, also detected in all the light curves. We executed the vetting module for this signal, which allowed us to rule out its systematic origin. We noticed that SPOC identified a threshold crossing event (TCE) for a candidate with an orbital period of 2.40\,d, that is, the first harmonic of the signal that \sherlock spotted. Hence, during the vetting stage, we carefully examined the depths of the odd/even eclipses and transit shapes in short and long cadences and found no remarkable difference among them. Moreover, the candidate did not show any problematic metric either on the transit source or the optical ghost checks, reinforcing our findings with a period of 1.20\,d as the more plausible solution. The TCE found by SPOC did not become a TOI due to its low S/N ($<$10) and slightly centroids offset. Next, we launched the statistical validation module, yielding values of FPP=0.18 and NFPP=0.00054, which fall on the likely planet area.  

Following the same procedure as in the case of the WASP-16 system, we conducted the Bayesian fitting and compared the Bayes Factor between two models: 1) only with the HAT-P-27\,b planet (null hypothesis) and 2) including the second potential planet found by \sherlock. We got $\Delta \ln Z = \Delta \ln Z_{2planet} - \Delta \ln Z_{1planet}\sim 64.6$, which strongly favors the 2-planet scenario over the simplest with a single planet. 
This candidate is then the CTOI-461239485.02, which corresponds to a hot Neptune-size planet of $\sim$4.3\,R$_{\oplus}$. Its parameters from the fitting can be consulted in Table~\ref{tab:fn_candidates}.  
The transit depth found for this candidate is $\sim$2.1\,ppt, which is deep enough for small- and medium-sized ground-based telescopes. Moreover, its orbital period is 1.20\,d, which eases scheduling and follow-up efforts due to the large number of possible observable events. However, the uncertainty in the transit time is $\Delta T\sim$ 3.3\,h, which is challenging to conduct time-critical observations from the ground but still possible. An alternative approach would be to conduct long observations from space-based telescopes such as CHEOPS, allowing for scanning up to 3$\sigma$ the predicted time without the day-night cycles, making a putative detection easier. All these considerations make CTOI-461239485.02 an excellent candidate to be confirmed or refuted with relatively low effort.

 The two-planet model is displayed in Fig.~\ref{fig:all_fit}, the posterior distribution is provided in Appendix~\ref{app:posterior_fitting}, Table~\ref{app:fits}, and the convergence of the model is visualized in Fig.~\ref{app:hat27_corner}. 

\begin{figure}
	\includegraphics[width=\columnwidth]{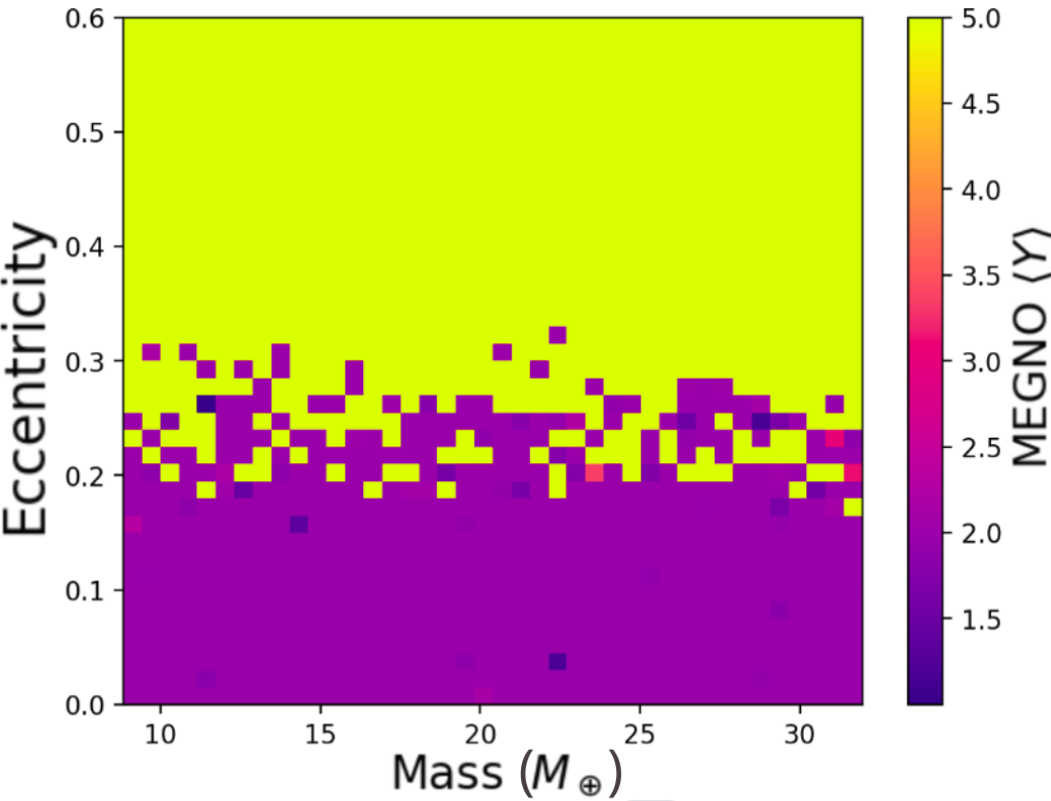}
    \caption{Stability map of the system HAT-P-27 composed of the two planets: HAT-P-27\,b and the planetary candidate CTOI-461239485.02. The map contains 40$\times$40 pixels, exploring the $e_{p}-M_{p}$
parameter space for the candidate. The stability of each scenario is evaluated using the MEGNO parameter, where $\langle Y(t) \rangle \rightarrow 2$ (purple shaded regions) represents quasi-periodic orbits and chaotic systems when $\langle Y(t) \rangle \rightarrow 5$ (yellow shaded regions).}
    \label{fig:hatp27_stability}
\end{figure}

Despite the vetting and validation tests passed by CTOI-461239485.02, the resulting planetary configuration with a hot Jupiter at 3.04\,d and an innermost candidate at 1.20\,d might be questionable due to stability concerns. For this reason, we executed the stability test (see Section~\ref{sec:megno}) assuming the nominal values for the star and HAT-P-27\,b planet as reported by \cite{beky2011}, and for the new candidate using a grid of 40 eccentricities from 0 to 0.6 and 40 planetary masses from 9.5 to 32\,$M_{\oplus}$ derived from the radius using a mass-radius relationship \citep[17$^{+13.8}_{-8.1}\,M_{\oplus}$;][]{chen2017}. This configuration allowed us to build a stability map exploring 1600 scenarios in the $e_{p}-M_{p}$ parameter space. Then, the MEGNO value is computed six times in each scenario by modifying randomly the orbital parameters. Hence, in total, the stability map explored 9600 scenarios. 
Each scenario was integrated over $\sim 10^{7}$ orbits of the outermost planet. The stability map in Fig.~\ref{fig:hatp27_stability} shows that the HAT-P-27 system can be stable for eccentricities up to 0.19 for the full range of masses explored for the planetary candidate. For larger eccentricities, the system becomes unstable for any of the masses within our explored grid. This analysis allowed us to conclude that despite the unusual architecture of the two-planet solution proposed here, consisting of a hot Jupiter and an inner hot Neptune, the system has ample room for stable configurations, resembling the WASP-84 system, where recently, an inner mini-Neptune has been found accompanying the previously known hot Jupiter WASP-84\,b \citep{wasp84}. 

\subsection{HAT-P-26}
HAT-P-26 is a K1 star with a $T_{eff}=5079\pm88\,K$, $R_{\star}=0.79_{-0.04}^{+0.10}\,R_{\odot}$ and $M_{\star}=0.82\pm0.03\,M_{\odot}$ observed nightly by the HAT-5 and HAT-6 telescopes of the HATNet project between January and August 2009, that hinted at the presence of a transiting planet with an orbital period of 4.23\,d. The planetary origin of this signal was later confirmed using the HIRES spectrograph on the Keck I Telescope from December 2009 to June 2010. All these observations together allowed the confirmation of the low-density Neptune-mass planet HAT-P-26\,b \citep{hartman2011} with a size of 0.565\,R$_{Jup}$, a mass of 0.059\,M$_{Jup}$ and an orbital period of 4.23\,d.

This star was observed by the TESS mission in Sector 50 in May 2022, using the 120 and 600\,sec cadences (TIC-420779000). During our search with \sherlock, we fully recovered the signal corresponding to HAT-P-26\,b with S/N=42.2 and SDE=15.6 in the first run. During the second run, \sherlock spotted an interesting signal 
with a period of 6.59\,d. While this signal was slightly below our detection threshold, with S/N=4.9 and SDE=7.4, it was confidently found in all the light curves. This motivated us to run the vetting module, where we did not find any solid potential systematic origin that explained this signal; however, it is important to note that due to its very low S/N, some of the tests were not conclusive. Still, we decided to conduct the statistical 
validation, which yielded FPP=0.24 and NFPP=0.0075, falling in the likely planet area.  

We then executed the fitting module to refine the planet parameters and compare the Bayes factor of the two models as for the other candidates. We found $\Delta \ln Z = \Delta \ln Z_{2planet} - \Delta \ln Z_{1planet}\sim 4.7$, which slightly favors the 2-planet scenario over the single-planet model. The candidate, named CTOI-420779000.02, would correspond to a mini-Neptune of $\sim$2.0\,R$_{\oplus}$ (see planet parameters in Table~\ref{tab:fn_candidates}), producing shallow transits of only $\sim$0.6\,ppt, which hinders any confirmation using ground-based telescopes. However, its confirmation would be feasible, as stated for the CTOI-46096489.02 in the WASP-16 system, by employing space-based telescopes such as CHEOPS. The uncertainty in the transit time is $\Delta T\sim$ 22.7\,h, which is too large for time-critical observations, and unfortunately, according to the TESS-point web tool, this star will not be revisited in future cycles, preventing us from reducing this uncertainty using upcoming TESS data. However, the orbital period of this candidate of only 6.59\,d would favor a filler follow-up camping to recover the planet by accumulating observations. This strategy is also applied in the FATE project using CHEOPS observations \citep{vangrootel2021} to find unknown transiting planets with orbital periods shorter than 10\,d. Following this strategy, it would be possible to confirm or refute the existence of this planet.  

The two-planet model is displayed in Fig.~\ref{fig:all_fit}, the posterior distribution is provided in Appendix~\ref{app:posterior_fitting}, Table~\ref{app:fits}, and the convergence of the model is visualized in Fig.~\ref{app:hat26_corner}.  

\begin{figure}
	\includegraphics[width=\columnwidth]{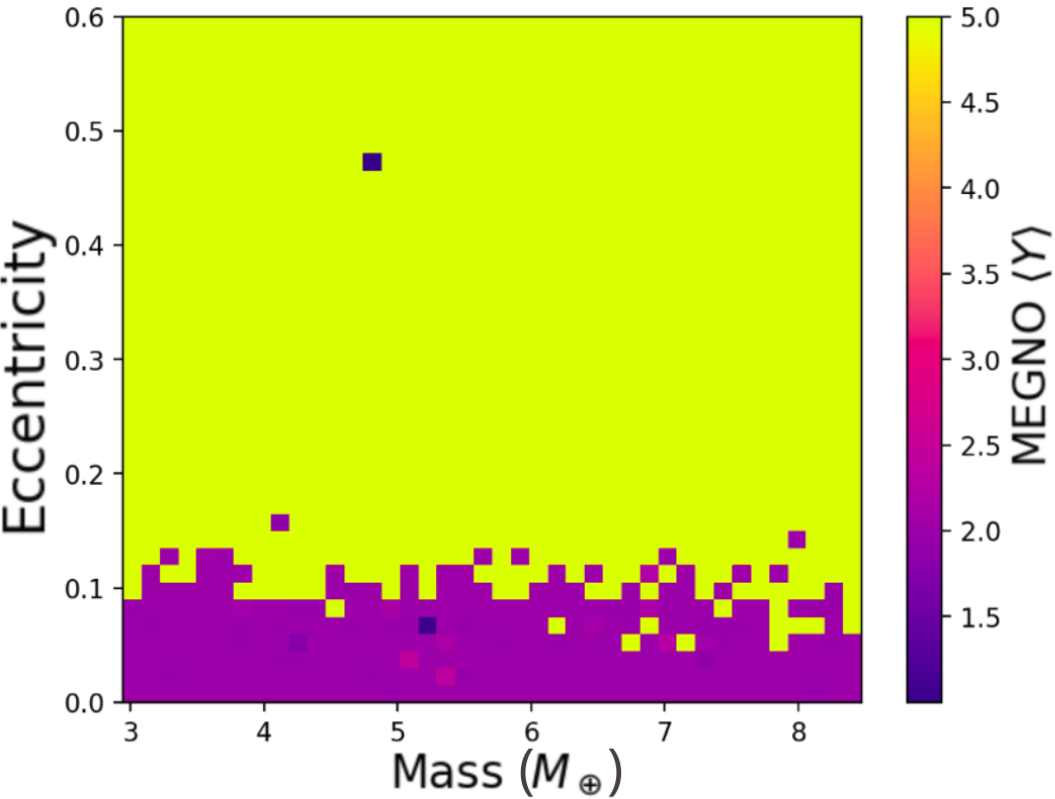}
    \caption{A stability map similar to the one displayed in Fig.~\ref{fig:hatp27_stability} but for the system HAT-P-26 composed of the two planets: HAT-P-26\,b and the planetary candidate CTOI-420779000.02.}
    \label{fig:hatp26_stability}
\end{figure}

As for HAT-P-27, the candidate's orbital period was close to the known planet HAT-P-26\,b. That architecture might be physically impossible and discarded by exploring the system stability. We, hence, followed the same strategy as for the HAT-P-27 system, building a stability map to explore 1600 scenarios in the $e_{p}-M_{p}$ parameter space, displayed in Fig.~\ref{fig:hatp26_stability}. In this analysis, the candidate's eccentricity must be lower than 0.05 to ensure the system's stability for all the planetary masses explored, while the scenarios with the candidate's mass below 6$\,M_{\oplus}$ show stable solutions for eccentricities up to 0.09 in most scenarios.

A particularly interesting point of the HAT-P-26 system is that \cite{essen2019} reported the detection of a drift in the spectroscopic data alongside a weak curvature in the timing residuals in the photometric data. The authors conducted a follow-up of the planet and obtained a $\sim$4\,min amplitude of transit timing variation (TTVs). While the authors prudently did not attribute these TTVs to a planetary companion, they gathered enough information to rule out the spot-induced origin of such TTVs. In a more recent study, \cite{thano2023} revisited the transit timings of HAT-P-26\,b over seven years, reporting a TTV amplitude of $\sim$2\,min, which might be attributable to the presence of an additional planet in the system of $\sim6.36\,M_{\oplus}$ with an orbital period of 8.47\,d, that is, in a first-order mean-motion resonance 2:1. However, there is not any hint of a transiting planet under this scenario. On the contrary, our solution points to the transiting nature of a potential planetary candidate of $\sim$1.96\,$R_{\oplus}$ (predicted mass from mass-radius relationships of 4.8$^{+3.5}_{-2.0}\,M_{\oplus}$) with an orbital period of 6.59\,d, that is, within $\sim$6$\%$ of the first-order resonance 3:2. The findings presented here can significantly encourage the scientific community to continue monitoring the HAT-P-26 system, to ultimately unravel its planetary architecture.

\subsection{TOI-2411}
TOI-2411 (TIC-10837041) is a K7-dwarf star with $T_{eff}=4099\pm123\,K$, $R_{\star}=0.68\pm0.02\,R_{\odot}$ and $M_{\star}=0.65\pm0.02\,M_{\odot}$. This star was monitored by 
the TESS mission in Sector 3 (October 2018) using 1800\,s cadence and in Sector 30 (October 2020) using 120 and 600\,s cadences. Orbiting TOI-2411 was alerted in November 2020 a planetary candidate, which was validated by \citep{giacalone2022} by combining high-resolution images collected using the SOAR/HRCam, Palomar/PHARO, and Keck/NIRC2 instruments, ground-based multi-band photometry (i' and r') taken employing the 1\,m Sinistro telescope at Las Cumbres Observatory and the 0.7\,m CDK telescope at the Mt. Kent Observatory, and statistical arguments. The planet, TOI-2411\,b, is a super-Earth of $1.68\pm0.11\,R_{\oplus}$ in an ultra-short period orbit of 0.78\,d. 

During our \sherlock search, we found in the first run a signal of 18.75\,d with a S/N=14.9 and SDE=14.1 in 9 of the 11 light curves processed. In the second run, we found a signal corresponding to TOI-2411\,b with S/N=8.4 and SDE=16.7 in all the light curves. We executed the vetting module for the 18.75\,d signal and found one red flag that might indicate a false positive: a momentum dump near the first transit. Momentum dumps might originate flux dimmings that can confuse the search algorithm; however, their shapes do not necessarily resemble transit-like features. We carefully visually inspected the two events in the 120\,s data, and they both look similar in shape, duration, and depth. Moreover, during the vetting, was also inspected the 600\,s cadence, and the signal corresponding to the 18.75\,d was also found. Hence, we concluded that the momentum dump, or any other systematic, was not the origin of the transit-like feature detected by \sherlock. Next, we conducted the statistical validation, which yielded values of FPP=0.022 and NFPP=0.0, falling in the likely planet area.

We then performed the fitting analysis and compared the two models as described for the previous candidates. We found $\Delta \ln Z = \Delta \ln Z_{2planet} - \Delta \ln Z_{1planet}\sim66.1$, which highly supports the two-planet model. The candidate, CTOI-10837041.02, is found to be a mini-Neptune of 2.88\,$R_{\oplus}$, whose transit depth is estimated to be $\sim$1.77\,ppt (see Table~\ref{tab:fn_candidates}), which can be detected using small- and medium size telescopes. However, its moderately long orbital period of 18.75\,d and estimated uncertainty in the transit time of $\Delta T\sim$ 6.7\,h make very challenging to confirm its planetary nature utilizing ground-based facilities. Unfortunately, TESS will not revisit this star, and hence, as in the case of CTOI-420779000.02 (HAT-P-26 system), a filler program using CHEOPS would be a potential solution to recover and confirm the candidate as a genuine planet.  

 The two-planet model is displayed in Fig.~\ref{fig:all_fit}, the posterior distribution is provided in Appendix~\ref{app:posterior_fitting}, Table~\ref{app:fits}, and the convergence of the model is visualized in Fig.~\ref{app:toi_corner}. 

\begin{figure*}
	\includegraphics[width=\textwidth]{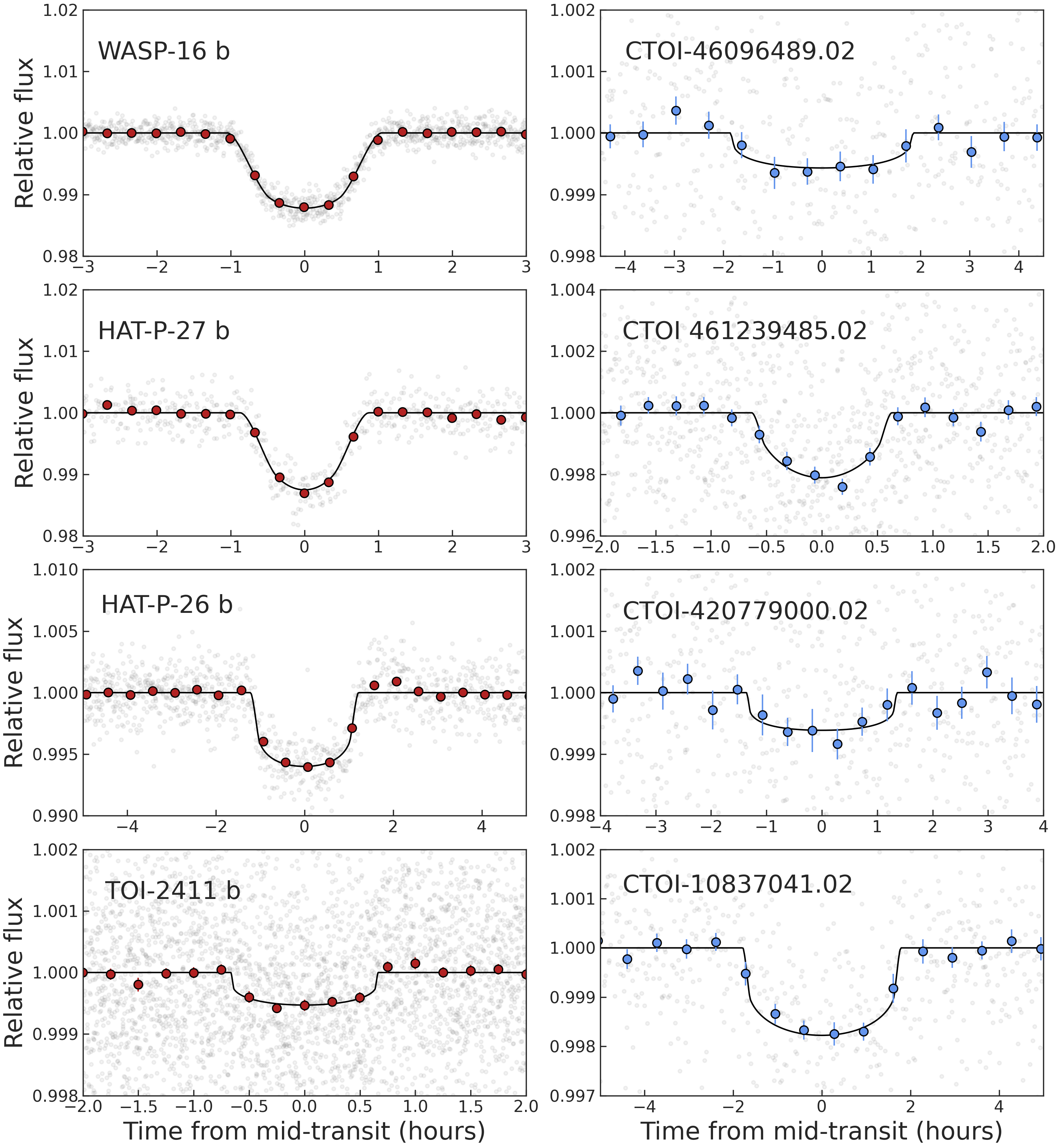}
    \caption{Detrended TESS photometry phase-folded to the periods of the known planets (left column) and planetary candidates (right columns), 
along with the best-fit transit models (solid black line). The unbinned data points are depicted in grey, while the colored circles with error bars correspond to binned data adapted in each case to ease the visualization. From top to bottom, the WASP-16, the HAT-P-27, the HAT-P-26, and the TOI-2411 systems. The posterior distributions
of these fits are provided in the Appendix~\ref{app:posterior_fitting} and the corner plots in Figs.~\ref{app:wasp16_corner},\ref{app:hat27_corner}, \ref{app:hat26_corner}, and \ref{app:toi_corner}, respectively.}
    \label{fig:all_fit}
\end{figure*}

\section{The Hidden Gems project}
\label{surveys}

Nearby and low-mass stars ($M<0.6\,M_\odot$) represent the best opportunity nowadays to study in great detail small transiting planets (R$_p$ $< 4$R$_{\oplus}$). Indeed, these stars ease the characterization of small exoplanets, not only in terms of radius and mass but also in terms of internal structure and atmospheric properties, which encodes information about physical and chemical properties, in turn providing insights into planets' formation and evolutionary history \citep[see, e.g.,][]{pudritz2018,triaud2021}. 

Summarizing, there are three critical points that support this reasoning: 1) their small sizes and lower masses compared to solar-like stars produce larger transit depths and induce larger radial velocities for a given planet, allowing for an accurate determination of planetary bulk properties, such as the radius and mass, with precisions better than $\sim$4$\%$ and $\sim$25$\%$, respectively. This allows us to derive the mean density and give a rough estimation of the bulk composition and internal structure \citep{otegi2020}. This advantage has been the driver of many transit and Doppler surveys such as MEarth \citep{nutzman2008}, SPECULOOS \citep{sebastian2021}, CARMENES \citep{quirrenbach2014}, and SPIRou \citep{artigau2014}, to name a few. 2) Atmospheres of planets orbiting these small stars can be explored using transit spectroscopy, which relies on the chromatic signature of the chemical species present in the atmosphere, detectable with current and future space-based observatories such as \textit{JWST} and \textit{Ariel}. Transit spectroscopy allows three configurations to study planetary atmospheres: Transmission spectroscopy, emission spectroscopy, and phase curves \citep{nikku2019}. The possibility of atmospheric characterization of small planets with any of these three techniques has been the drive behind the detection of new planets around such stars, notably thanks to the \textit{TESS} and \textit{Kepler/K2} missions. It has even now become one of the main science goals of the community, with its importance recognized at the highest level with a recommendation by the \textit{JWST} Exoplanet Working Group to dedicate over 500 hours of Director's Discretionary Time to this end\footnote{\url{https://www.stsci.edu/files/live/sites/www/files/home/hst/about/space-telescope-users-committee/presentations-and-documentation/_documents/2023_nov/exoplanets-DDWG.pdf}}. 3) Low-mass stars are the most common in the solar neighborhood \citep{2015_Winters_Mdwarfs_solar_neighborhood}, representing up to 70$\%$ of the stellar content of our Galaxy. Moreover, they have a lower binary fraction than solar-like stars. In addition, computational simulations, as well as Doppler and transit surveys, have concluded that the fraction of stars hosting small planets increases with decreasing mass \citep[see, e.g.,][]{alibert_formation_2017,bonfils2013,2015_occurence_rates}. By combining the high frequency of low-mass stars with the occurrence rate of small planets orbiting them, there might exist $\sim$10$\times$ more small planets orbiting these stars than around Solar analogs. Hence, planets discovered orbiting low-mass stars are more representative of planet formation processes than planets orbiting single Sun-like stars \citep{triaud2021}.

Of particular interest are these systems hosted by a low-mass star where more than one transiting planet exists. In these cases, by studying different planets that were born in the same protoplanetary disc, we can get information not only about the planets but also about the formation and evolution of the whole planetary system. By aggregating information gleaned from the analysis of numerous multiplanetary systems, we will be able to formulate a robust planetary formation model that can resolve existing degeneracies and overcome the limitations currently constraining our understanding \citep[see, e.g.,][]{raymond2022}. This comprehensive model will eventually enable us to systematically decipher the underlying processes that govern planetary formation and evolution across different environments, unifying disparate observations and refining our predictions regarding the distribution and characteristics of exoplanets.

The full sample of transiting exoplanets around low-mass stars keeps on growing and now exceeds 1400\footnote{NASA Exoplanet Archive (20 Feb 2024) \url{https://exoplanetarchive.ipac.caltech.edu/}} planets with 85$\%$ of them being smaller than Neptune. As the race to detect new small planets amenable for detailed studies is still ongoing, we present the Hidden Gems project, which relies on the capabilities of \sherlock to discover extra planets in low-mass star systems hosting at least one transiting planet. 

\begin{figure}
	\includegraphics[width=\columnwidth]{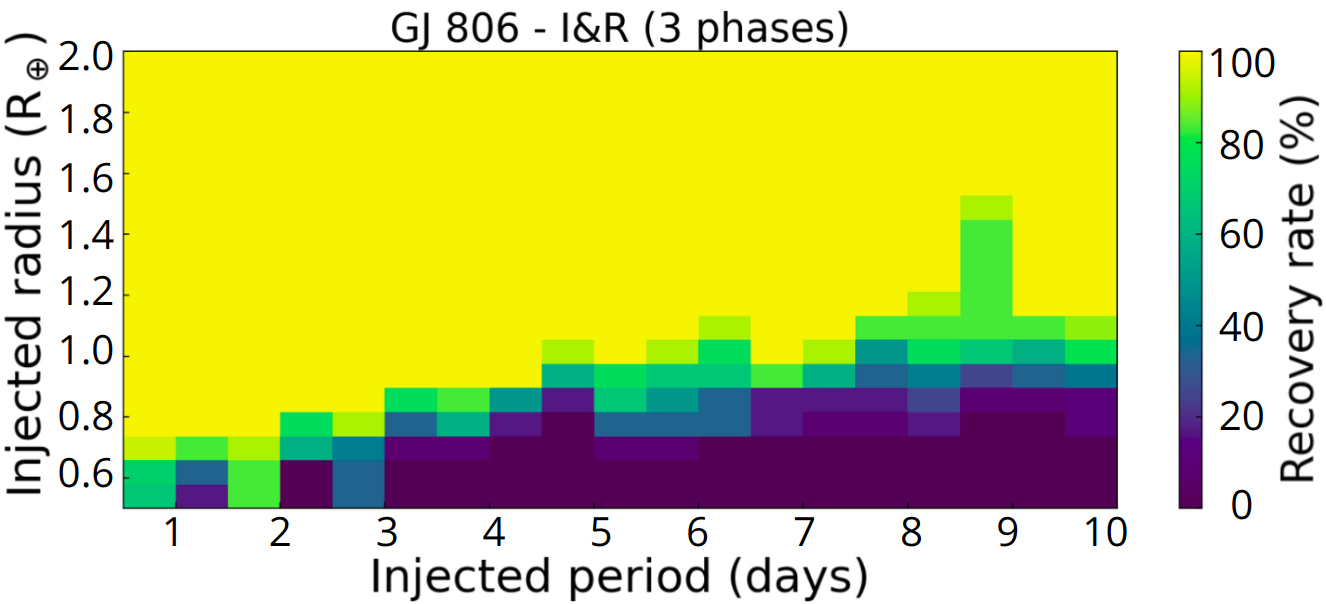}
	\includegraphics[width=\columnwidth]{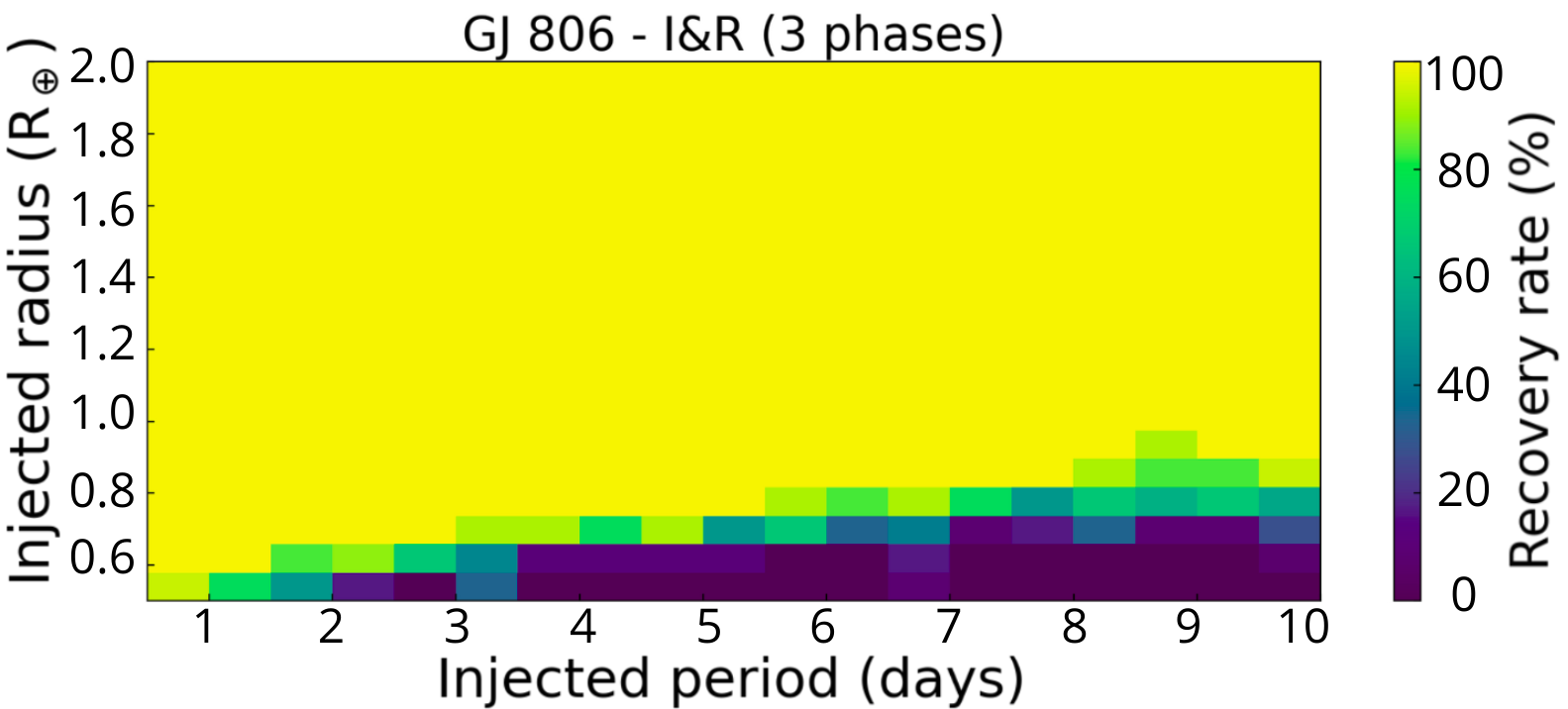}
    \caption{Improvements between the injection and recovery scenarios for a simple search on GJ 806 using different number of sectors. Above, results of injection and recovery scenarios only using sector 41. Bottom, same retrieval scenarios using sectors 41, 55 and 56.}
    \label{fig:inj-rec-gj806}
\end{figure}

Our target list is constructed by filtering stars with T$_{\mathrm{eff}} < 5300$\,K to include K- and M-dwarf stars and select a volume-limited sample within a distance of 50 pc, which totals 167 transiting planets in 107 unique systems at the time of writing. However, this catalog will be updated with the upcoming releases of newly discovered planetary systems. Occurrence rate studies show that M dwarfs host at least 2 planets on average \citep[e.g.][]{2015_occurence_rates,tuomi2019frequency}, this corresponds to twice the number reported by \cite{2020AJ_K_occurence_rate} for K dwarfs. It implies, on a statistical level, that there could still be additional planets in these systems. Moreover, a transiting system has a favorable orbital inclination that enhances the likelihood of finding extra planet candidates. \textit{TESS} offers an exquisite opportunity, thanks to its long baselines of observations, to search for concealed low S/N signals that might have been unnoticed. The search will be conducted on each system's available \textit{TESS} sectors and will be updated as soon as new data becomes public. This strategy increases the S/N of the putative detections, reaching smaller planetary radii and longer period candidates, as shown in Fig.~\ref{fig:inj-rec-gj806}. 

By developing this project, we will eventually be able to pinpoint the presence of small transiting planets in the habitable zones of their stars within the solar neighborhood, setting the stage for detailed studies. Hence, the ultimate objective of this project is to compile an accurate census of nearby transiting potential habitable planets.

\section{Conclusions}
\label{conclusions}
Providing open-sourced software packages is a valuable contribution to the scientific community, especially in cases where the documentation is abundant, detailed, and focused on hands-on examples. With \sherlock, we are not only doing that but also releasing a mature pipeline standardizing scientific procedures that increase the scientist's productivity, help the reproducibility of the end-to-end processes, and empower individual teams to conduct their own planetary searches relying on space-based missions.

\sherlock represents an advanced exoplanetary search pipeline that leverages a suite of robust and community-validated astrophysical packages including \lightkurve, \eleanor, \tls, \triceratops, and \allesfitter. These are complemented by bespoke modules tailored to create a streamlined and integrated workflow. Initially, \sherlock systematically acquires and preprocesses available data, setting the stage for effective exoplanet detection. Subsequent phases involve rigorous vetting of candidates to identify and eliminate potential false positives through comprehensive analysis and statistical validation techniques. The pipeline further refines estimates of planetary characteristics through sophisticated data modeling, enhancing the precision of the derived orbital and physical parameters. \sherlock concludes the analyses by determining the best observational windows, crucial for planning ground-based confirmatory follow-ups. Each phase is encapsulated within dedicated modules, optimizing the accuracy and efficiency of the whole process of finding reliable planetary candidates. Still, we highlight that \sherlock may need manual fine-tuning to operate to its fullest potential, making it not an entire automated search pipeline, requiring visual inspection of detected signals before passing through all the modules.      

Our performance test showed that \sherlock has an excellent recall of 0.98 when recovering SPOC official alerts confirmed as planets, which makes it a very reliable pipeline for the community. During this test, we not only recovered 98$\%$ of the planets but also found some candidates that were confirmed planets but did not have any associated alerts, such as TOI-942\,c and Kepler-89 c, strengthening the conclusion of being a reliable planetary searching tool. On top of that, we found four new planetary candidates unnoticed until now that fulfilled all our tests, only lacking ground-based observation to confirm their planetary nature fully. These four candidates were found to be in the planetary systems WASP-16 (CTOI-46096489.02), HAT-P-27 (CTOI-461239485.02), HAT-P-26 (CTOI-420779000.02) and TOI-2411 (CTOI-10837041.02).     

The vetting implemented by \sherlock, based on the robust tests conducted by the SPOC pipeline, reports a variety of metrics that are marked as red flags whenever they are found to be above specific thresholds. However, low data quality sometimes produces inconclusive results for these metrics, especially on the centroids and optical ghost diagnoses. Hence, in these cases, the red flags have to be taken as warning bells instead of a robust indication of a false positive, which needs further exploration by the user.   
\sherlock's vetting will be further developed in coming releases towards a more robust data exploration, producing a more reliable set of metrics even for very low S/N signals. In particular, we are currently exploring implementing a deep learning neural network similar to ExoMiner \citep{valizadegan2022}, allowing us to discern genuine planets from false positives unbiasedly.

\section*{Acknowledgements}
We thank the anonymous referee for the helpful comments. F.J.P. acknowledges financial support from the Severo Ochoa grant CEX2021-001131-S funded by MCIN/AEI/10.13039/501100011033 and through the project PID2022-137241NB-C43. 
We acknowledge support from the Spanish Ministry of Science through the project PID2019-107061GB-C64/SRA (State Research Agency/10.13039/501100011033). Funding for open access charge: Universidad de Granada / CBUA. This publication benefits from the support of the French Community of Belgium in the context of the FRIA Doctoral Grant awarded to M.T. V.V.G. is an F.R.S.-FNRS Research Associate.

\section*{Data Availability}

The data underlying this article will be shared on reasonable request to the corresponding author.



\bibliographystyle{mnras}
\bibliography{example} 




\appendix
\section{TIC 169285097 \sspe fit signals}
\label{app:sspe_fit_signals}
\begin{figure*}
	\includegraphics[width=0.33\textwidth]{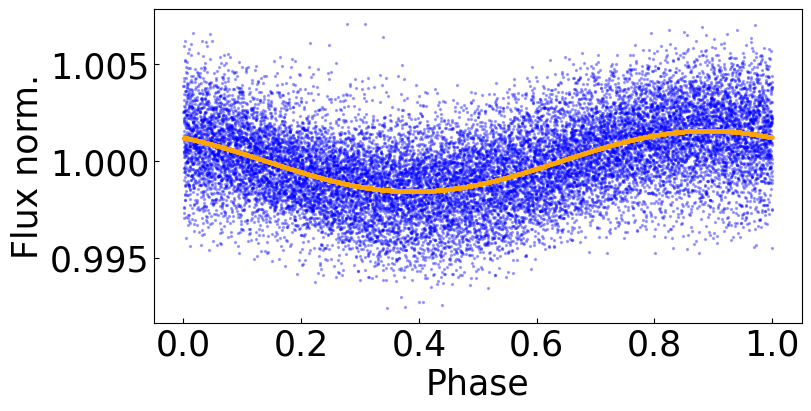}
	\includegraphics[width=0.33\textwidth]{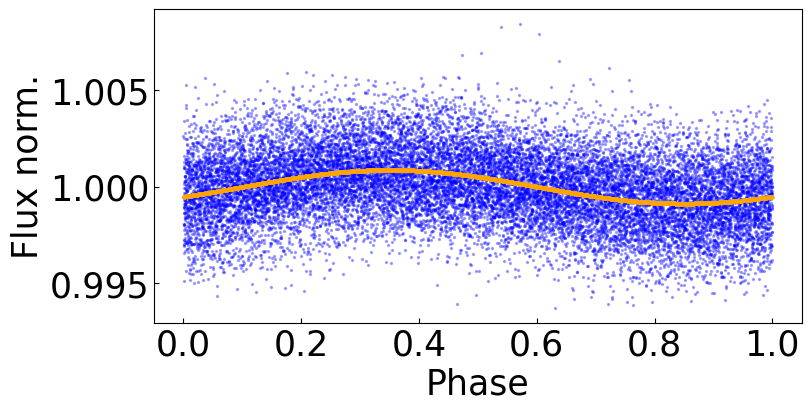}
	\includegraphics[width=0.33\textwidth]{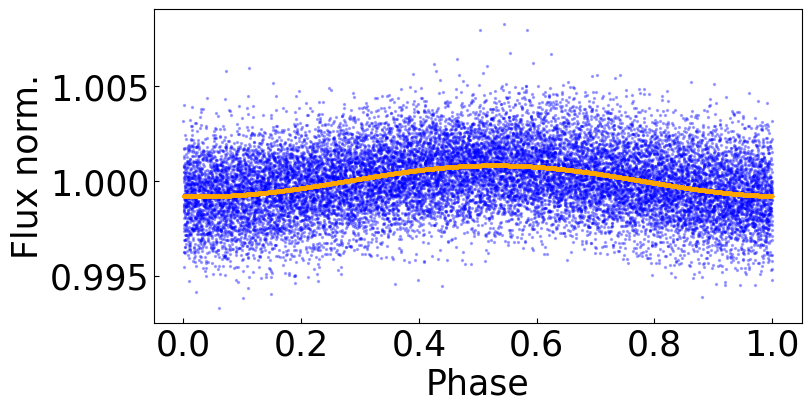}
	\includegraphics[width=0.33\textwidth]{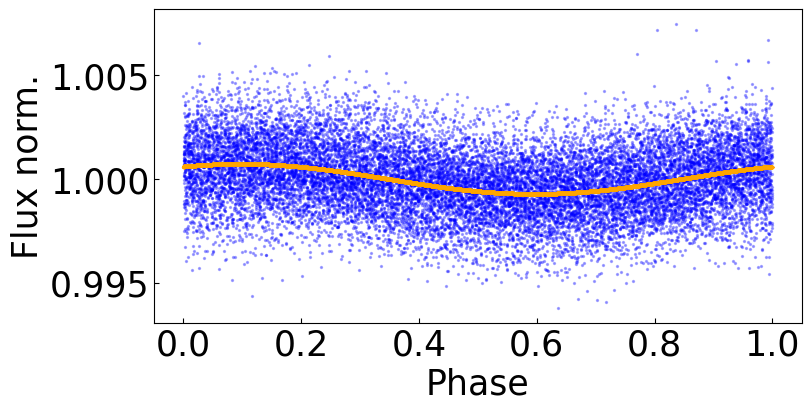}
	\includegraphics[width=0.33\textwidth]{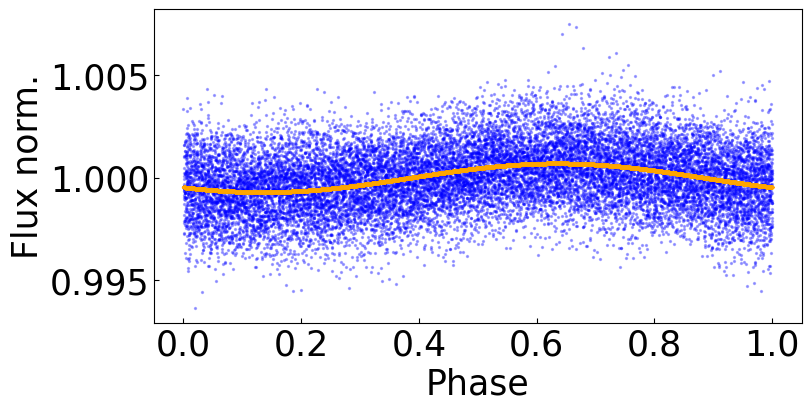}
	\includegraphics[width=0.33\textwidth]{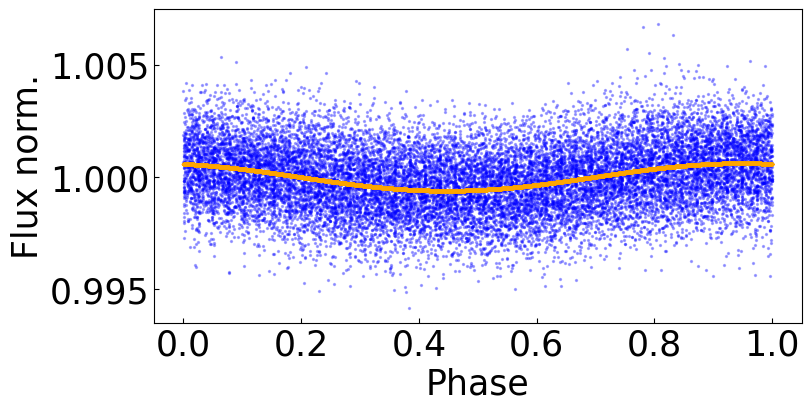}
	\includegraphics[width=0.33\textwidth]{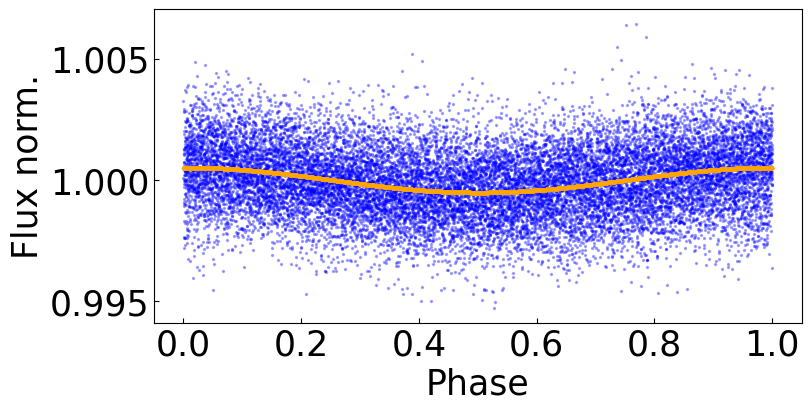}
	\includegraphics[width=0.33\textwidth]{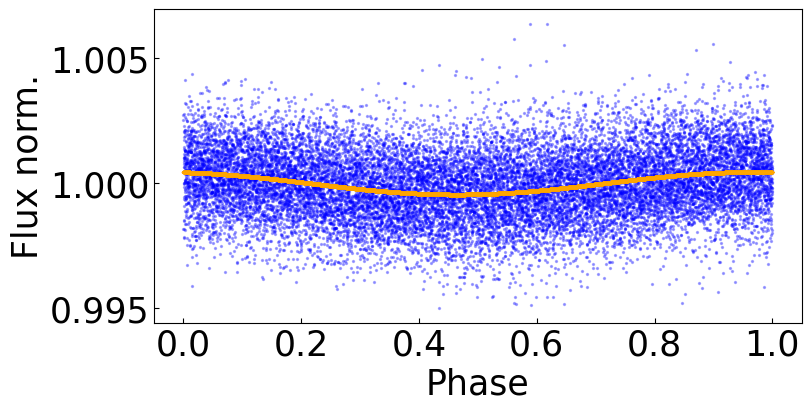}
	\includegraphics[width=0.33\textwidth]{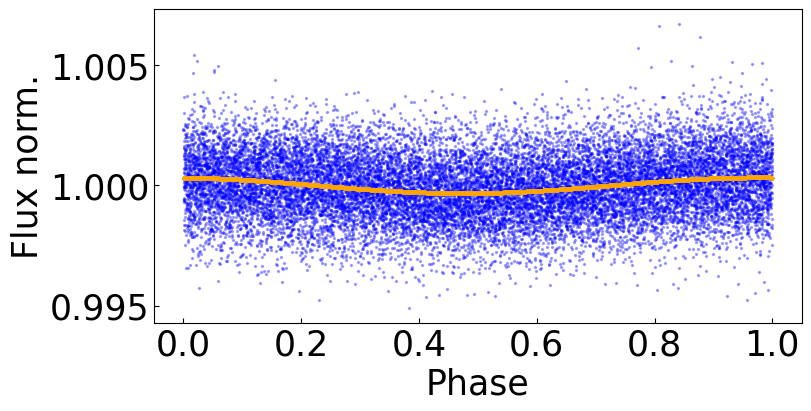}
    \caption{Output from \sherlock \sspe algorithm for Sector 2 data of TIC 169285097. From top-left to bottom-right, the first nine pulsations fits are represented after the previous ones have already been subtracted. The blue dots show the phased flux, and the orange lines show the best sinusoidal fit.}
    \label{fig:pulsation-modes}
\end{figure*}

\section{Performance survey catalog: TIC ids}
\label{app:recall_catalogue}
100566492, 102264230, 10837041, 111991770, 112099249, 120247528, 12862099, 136916387, 139375960, 142937186, 144700903, 146520535, 147797743, 157230659, 165297570, 168281028, 172598832, 175772482, 181949561, 188876052, 189380158, 189625051, 20096620, 201186294, 201604954, 206284139, 207077681, 21113347, 229047362, 237320326, 243921117, 245392284, 248853232, 250707118, 257567854, 262715204, 263179590, 269217040, 27039476, 273231214, 281408474, 283419843, 288246496, 291685334, 291751373, 292152376, 305048087, 32487566, 33595516, 336732616, 339522221, 347430350, 354619337, 368739874, 368805700, 369455629, 372172128, 374530847, 375942197, 37718056, 381856447, 391958006, 392476080, 398895470, 39926974, 399402994, 401604346, 420779000, 420814525, 425817867, 429302040, 431701493, 434226736, 436478932, 437704321, 441420236, 452808876, 453064665, 459730973, 460396820, 46096489, 461239485, 464646604, 466382581, 466390120, 467971286, 49043968, 51637609, 53728859, 54002556, 57984377, 6663331, 7088246, 90850770, 91051152, 9155187, 92449173, 9385460, 9443323, 97409519

\section{Posterior distributions from the fits}
\label{app:posterior_fitting}

\begin{table*}
    \centering
    \caption{Posterior distributions for WASP-16, HAT-P-27, HAT-P-26 and TOI-2411 systems}
    \label{app:fits}
    \begin{tabular}{lccc} 
     \hline 
     Parameter & Unit & WASP-16\,b & CTOI-46096489.02 \\ 
     \hline 
     \noalign{\smallskip}
     $R_p / R_\star$ & & $0.1150_{-0.0019}^{+0.0022}$ & $0.0215_{-0.0022}^{+0.0017}$ \\ \noalign{\smallskip}
     $(R_\star + R_p) / a_p$ & & $0.1266_{-0.0040}^{+0.0036}$ & $0.0493_{-0.0031}^{+0.0040}$ \\ \noalign{\smallskip}
     $\cos{i_p}$ & & $0.0921_{-0.0051}^{+0.0046}$ & $0.016_{-0.010}^{+0.013}$  \\ \noalign{\smallskip}
     Mid-transit time, $T_{0;p}$ & $\mathrm{BJD}_{TDB} - 2457000$  & $1613.66548\pm0.00020$ & $1618.725_{-0.014}^{+0.012}$ \\ \noalign{\smallskip}
     Orbital Period, $P_p$ & $\mathrm{days}$ & $3.118531\pm0.000073$ &  $10.457_{-0.028}^{+0.018}$ \\ \noalign{\smallskip}
     $q_{1}$ & & \multicolumn{2}{c}{$0.60_{-0.20}^{+0.23}$} \\ \noalign{\smallskip}
     $q_{2}$ & & \multicolumn{2}{c}{$0.35_{-0.21}^{+0.28}$} \\ \noalign{\smallskip}
     $\log{\sigma_{w}}$ & $\log{ \mathrm{rel. flux.} }$ & \multicolumn{2}{c}{$-6.5560\pm0.0055$}  \\ \noalign{\smallskip}
     $\mathrm{GP: \ln{\sigma}}$ & & \multicolumn{2}{c}{$-8.69_{-0.15}^{+0.16}$} \\ \noalign{\smallskip}
     $\mathrm{GP: \ln{\rho}}$ & & \multicolumn{2}{c}{$-0.75_{-0.31}^{+0.34}$} \\ \noalign{\smallskip}
     \hline 
     Parameter & Unit & HAT-P-27\,b & CTOI-461239485.02 \\ 
     \hline 
     \noalign{\smallskip}
     $R_p / R_\star$ & & $0.1209_{-0.0045}^{+0.0050}$ & $0.0457_{-0.0040}^{+0.0042}$ \\ \noalign{\smallskip}
     $(R_\star + R_p) / a_p$ & & $0.1069_{-0.0063}^{+0.0059}$ & $0.188_{-0.017}^{+0.025}$ \\ \noalign{\smallskip}
     $\cos{i_p}$ & & $0.0767\pm0.0076$ & $0.0130_{-0.031}^{+0.033}$  \\ \noalign{\smallskip}
     Mid-transit time, $T_{0;p}$ & $\mathrm{BJD}_{TDB} - 2457000$  & $2705.8727\pm0.0004$ & $2705.5090_{-0.0015}^{+0.0014}$ \\ \noalign{\smallskip}
     Orbital Period, $P_p$ & $\mathrm{days}$ & $3.03908\pm0.000019$ &  $1.19949_{-0.00022}^{+0.00021}$ \\ \noalign{\smallskip}
     $q_{1}$ & & \multicolumn{2}{c}{$0.75_{-0.22}^{+0.16}$} \\ \noalign{\smallskip}
     $q_{2}$ & & \multicolumn{2}{c}{$0.67_{-0.33}^{+0.21}$} \\ \noalign{\smallskip}
     $\log{\sigma_{w}}$ & $\log{ \mathrm{rel. flux.} }$ & \multicolumn{2}{c}{$-6.0260\pm0.0073$}  \\ \noalign{\smallskip}
     $\mathrm{GP: \ln{\sigma}}$ & & \multicolumn{2}{c}{$-6.83_{-0.24}^{+0.28}$} \\ \noalign{\smallskip}
     $\mathrm{GP: \ln{\rho}}$ & & \multicolumn{2}{c}{$0.52\pm0.31$} \\ \noalign{\smallskip}
     \hline 
     Parameter & Unit & HAT-P-26\,b & CTOI-420779000.02 \\ 
     \hline 
     \noalign{\smallskip}
     $R_p / R_\star$ & & $0.0722_{-0.0016}^{+0.0015}$ & $0.0230\pm0.0024$ \\ \noalign{\smallskip}
     $(R_\star + R_p) / a_p$ & & $0.0821_{-0.0013}^{+0.0014}$ & $0.0584\pm0.0010$ \\ \noalign{\smallskip}
     $\cos{i_p}$ & & $0.0331_{-0.0049}^{+0.0042}$ & $0.022\pm0.014$  \\ \noalign{\smallskip}
     Mid-transit time, $T_{0;p}$ & $\mathrm{BJD}_{TDB} - 2457000$  & $2678.8923\pm0.0005$ & $2683.1598_{-0.0014}^{+0.0010}$ \\ \noalign{\smallskip}
     Orbital Period, $P_p$ & $\mathrm{days}$ & $4.23440\pm0.00031$ &  $6.5938_{-0.0069}^{+0.0087}$ \\ \noalign{\smallskip}
     $q_{1}$ & & \multicolumn{2}{c}{$0.44_{-0.23}^{+0.34}$} \\ \noalign{\smallskip}
     $q_{2}$ & & \multicolumn{2}{c}{$0.26_{-0.16}^{+0.31}$} \\ \noalign{\smallskip}
     $\log{\sigma_{w}}$ & $\log{ \mathrm{rel. flux.} }$ & \multicolumn{2}{c}{$-6.3901\pm0.0064$}  \\ \noalign{\smallskip}
     $\mathrm{GP: \ln{\sigma}}$ & & \multicolumn{2}{c}{$-8.78_{-0.23}^{+0.19}$} \\ \noalign{\smallskip}
     $\mathrm{GP: \ln{\rho}}$ & & \multicolumn{2}{c}{$-1.92_{-0.75}^{+1.3}$} \\ \noalign{\smallskip}
     \hline 
     Parameter & Unit & TOI-2411\,b & CTOI-10837041.02 \\ 
     \hline 
     \noalign{\smallskip}
     $R_p / R_\star$ & & $0.021\pm0.001$ & $0.0390_{-0.0018}^{+0.0016}$ \\ \noalign{\smallskip}
     $(R_\star + R_p) / a_p$ & & $0.2230_{-0.0055}^{+0.0060}$ & $0.02741_{-0.00081}^{+0.00094}$ \\ \noalign{\smallskip}
     $\cos{i_p}$ & & $0.0334_{-0.020}^{+0.017}$ & $0.0115_{-0.0030}^{+0.0024}$  \\ \noalign{\smallskip}
     Mid-transit time, $T_{0;p}$ & $\mathrm{BJD}_{TDB} - 2457000$  & $2129.3198_{-0.0016}^{+0.0012}$ & $2141.3147_{-0.0034}^{+0.0034}$ \\ \noalign{\smallskip}
     Orbital Period, $P_p$ & $\mathrm{days}$ & $0.78289_{-0.00018}^{+0.00014}$ &  $18.7496_{-0.0041}^{+0.0047}$ \\ \noalign{\smallskip}
     $q_{1}$ & & \multicolumn{2}{c}{$0.41_{-0.23}^{+0.34}$} \\ \noalign{\smallskip}
     $q_{2}$ & & \multicolumn{2}{c}{$0.41_{-0.25}^{+0.32}$} \\ \noalign{\smallskip}
     $\log{\sigma_{w}}$ & $\log{ \mathrm{rel. flux.} }$ & \multicolumn{2}{c}{$-6.6475\pm0.0045$}  \\ \noalign{\smallskip}
     $\mathrm{GP: \ln{\sigma}}$ & & \multicolumn{2}{c}{$-8.961\pm0.089$} \\ \noalign{\smallskip}
     $\mathrm{GP: \ln{\rho}}$ & & \multicolumn{2}{c}{$-1.49_{-0.35}^{+0.40}$} \\ \noalign{\smallskip}
     \hline 
     
     \end{tabular}
     
\end{table*}


\begin{figure*}
	\includegraphics[width=\textwidth]{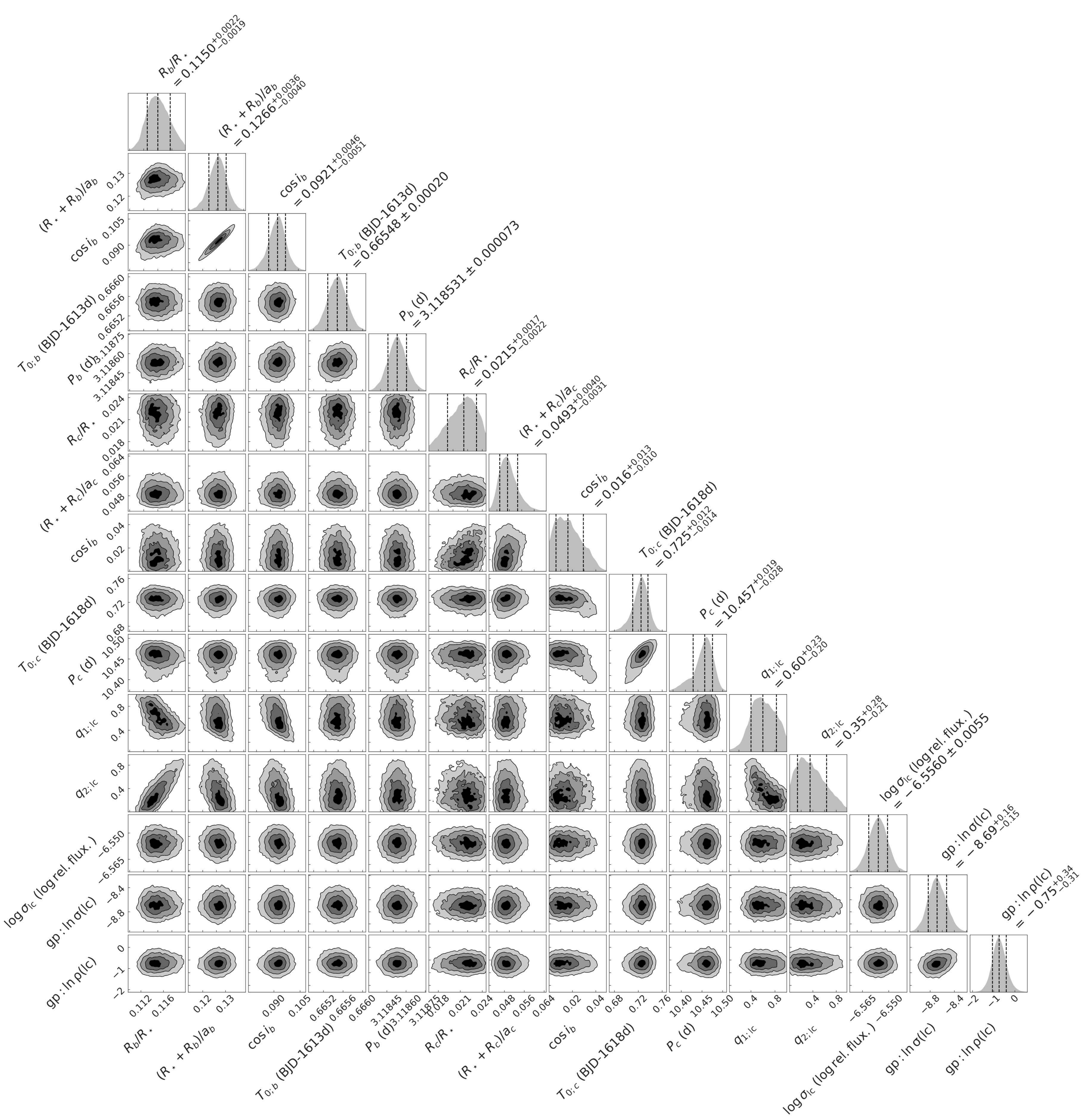}
    \caption{Posterior probability distributions for the fitting parameters used to model the planetary system WASP-16 as described in Section~\ref{sec:bayes} considering the two-planet scenario from Section~\ref{sec:wasp16}. The vertical dashed lines represent the median and the 68$\%$ credible interval. The figure highlights the correlation (or absence thereof) between all the parameters. The values of each parameter can be consulted in Table~\ref{app:fits}.}
    \label{app:wasp16_corner}
\end{figure*}

\label{app:hat27}
\begin{figure*}
	\includegraphics[width=\textwidth]{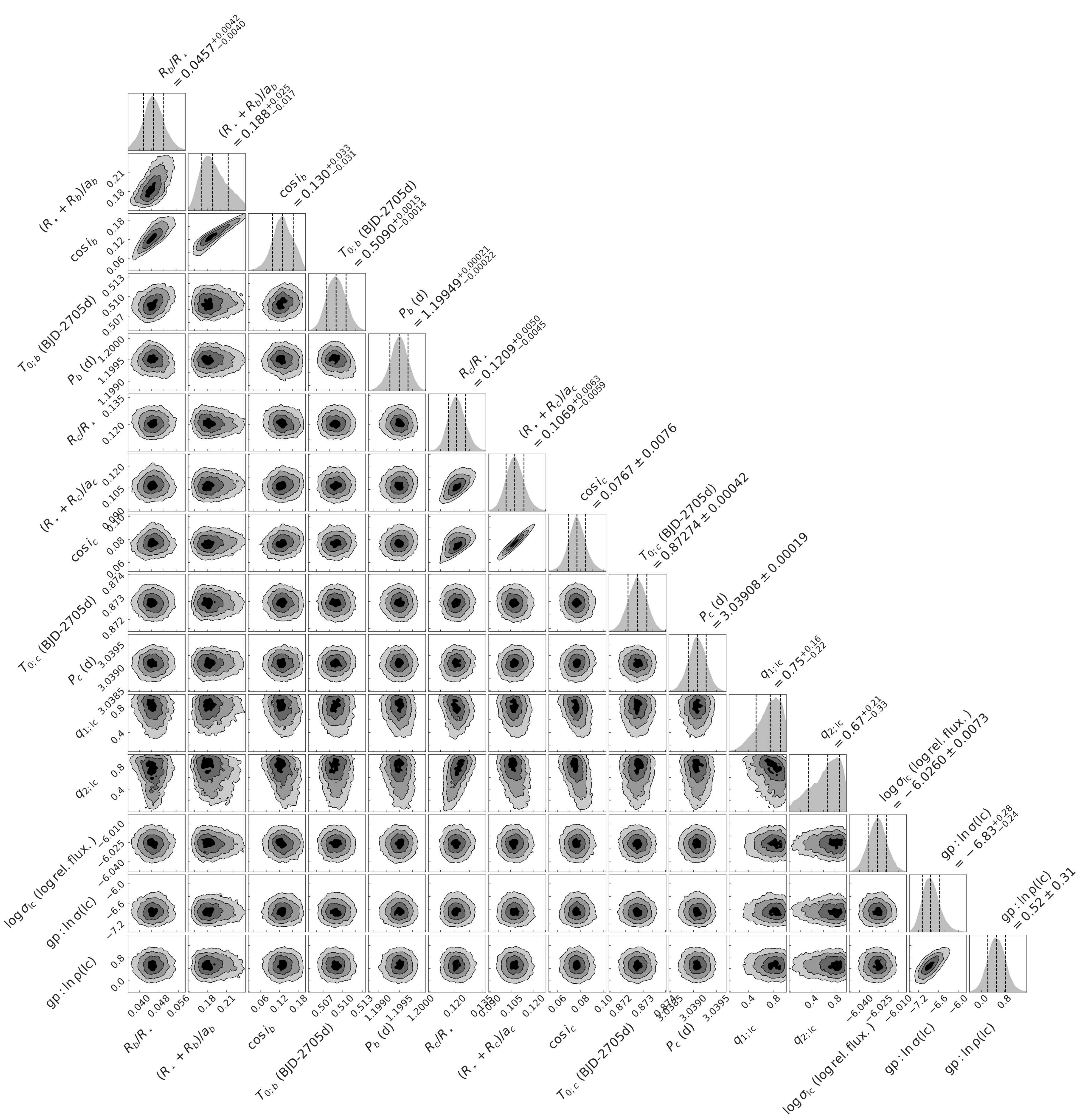}
    \caption{Similar to Fig.~\ref{app:wasp16_corner}, but for the HAT-P-27 system.}
    \label{app:hat27_corner}
\end{figure*}

\label{app:hat26}
\begin{figure*}
	\includegraphics[width=\textwidth]{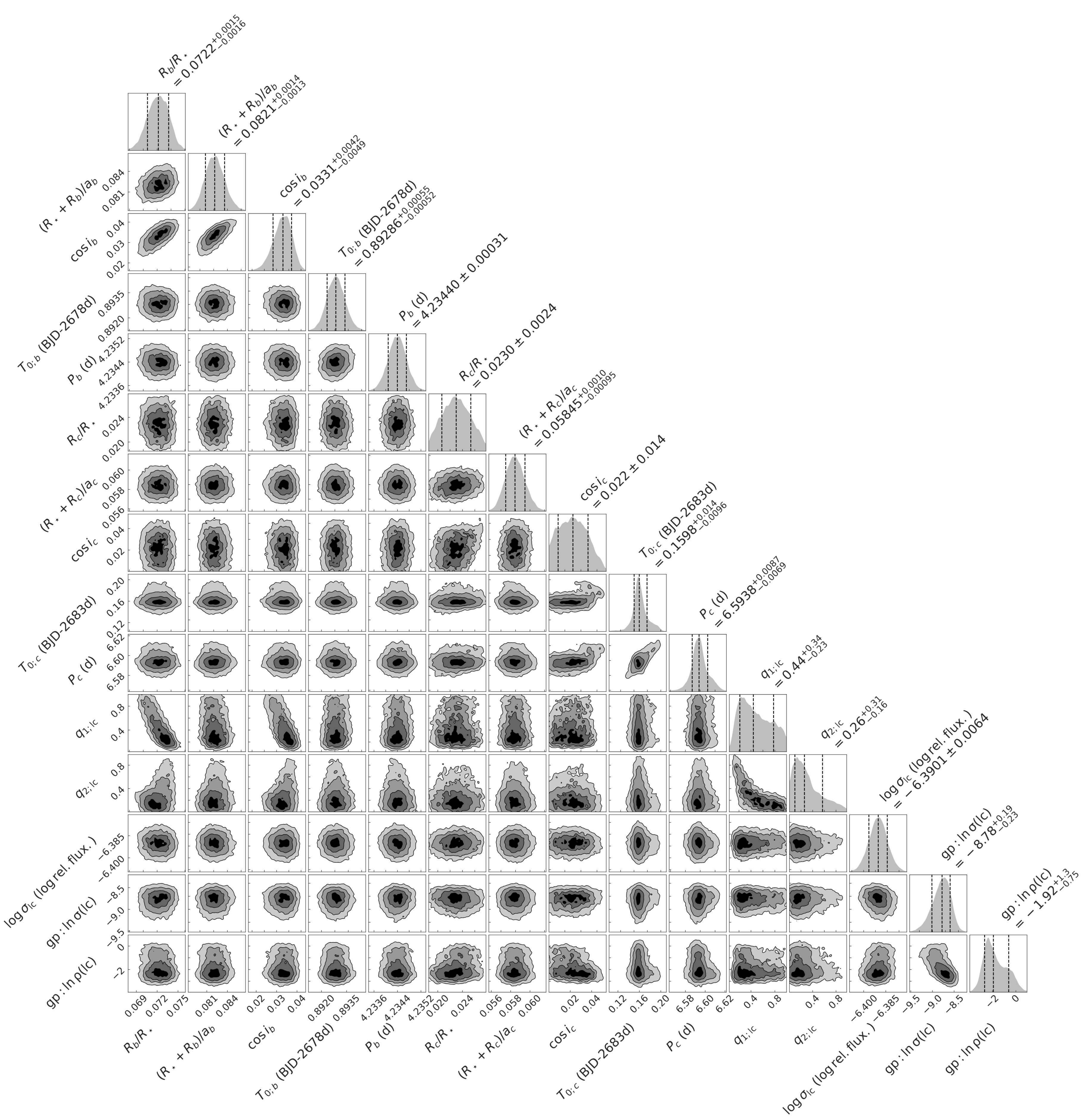}
    \caption{Similar to Fig.~\ref{app:wasp16_corner}, but for the HAT-P-26 system.}
    \label{app:hat26_corner}
\end{figure*}

\label{app:toi}
\begin{figure*}
	\includegraphics[width=\textwidth]{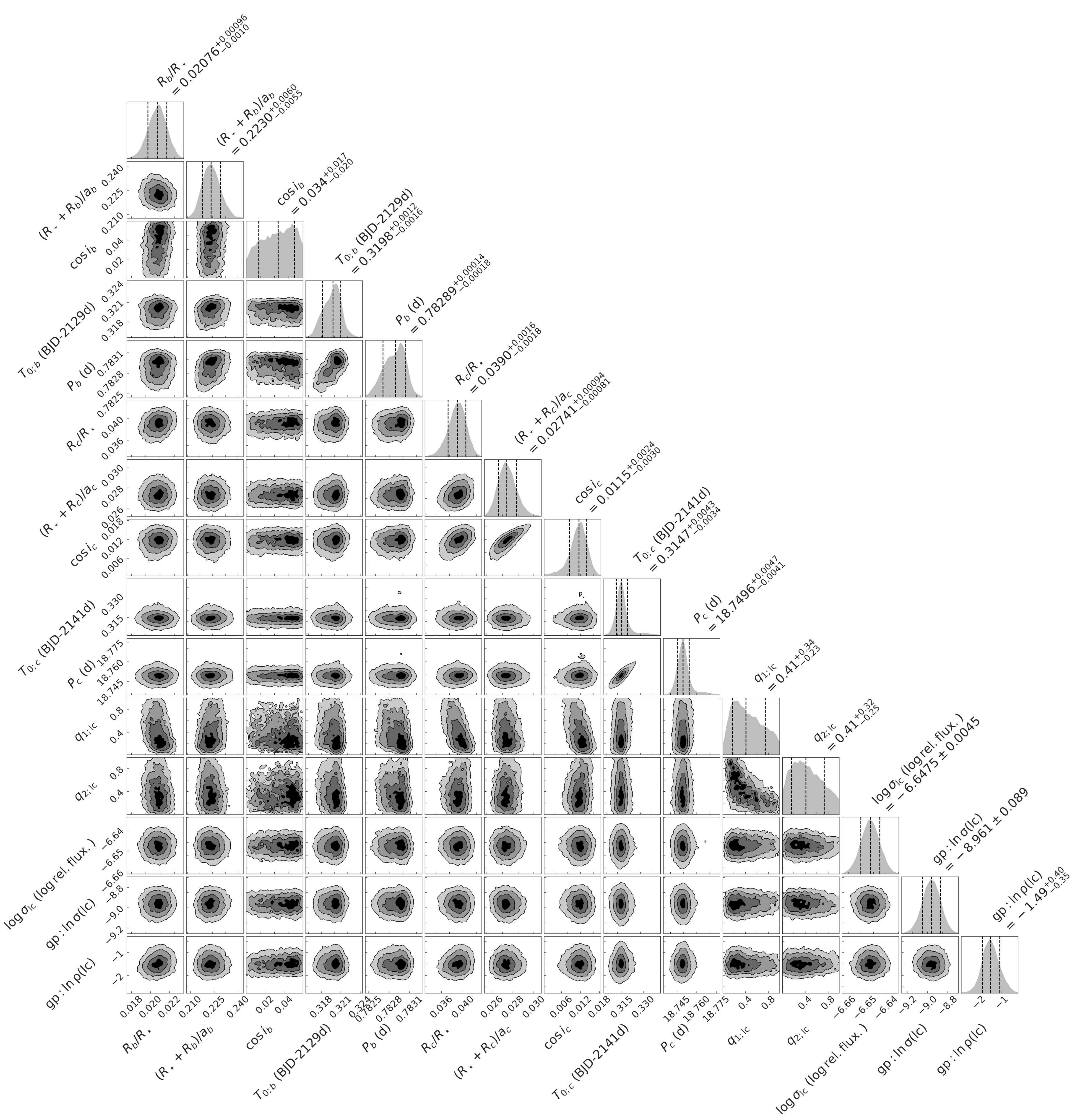}
    \caption{Similar to Fig.~\ref{app:wasp16_corner}, but for the TOI-2441 system.}
    \label{app:toi_corner}
\end{figure*}

\bsp	
\label{lastpage}
\end{document}